\documentclass[fleqn,usenatbib]{mnras}
\usepackage{newtxtext,newtxmath}
\usepackage{comment}
\usepackage[T1]{fontenc}
\usepackage{placeins}
\usepackage{bm}

\DeclareRobustCommand{\VAN}[3]{#2}
\let\VANthebibliography\thebibliography
\def\thebibliography{\DeclareRobustCommand{\VAN}[3]{##3}\VANthebibliography}

\newcommand{\rtext}[1]{#1}
\newcommand{\ptext}[1]{#1}

\usepackage{graphicx}	
\usepackage{amsmath}	

\title[Dark Matter Substructure or Source Model Systematics?]{Dark Matter Substructure or Source Model Systematics? \\ A Case Study of Cluster Lens Abell S1063}

\author[Ephremidze et al.]{
Nino Ephremidze,$^{1}$\thanks{E-mail: nino\_ephremidze@college.harvard.edu}
Chandrika Chandrashekar,$^{1}$\thanks{E-mail:
cchandrashekar@g.harvard.edu}
Atınç Çağan Şengül$^{2}$\thanks{E-mail:
aa2@pitt.edu},
and Cora Dvorkin$^{1}$\thanks{E-mail: cdvorkin@g.harvard.edu}
\\
$^{1}$Department of Physics, Harvard University, Cambridge, MA 02138, USA\\
$^{2}$Pittsburgh Particle Physics, Astrophysics, and Cosmology Center (PITT PACC),
Department of Physics and Astronomy, University of Pittsburgh,\\
3941 O'Hara Street, Pittsburgh, PA, 15260
}

\pubyear{\the\year{}}

\begin{document}
\label{firstpage}
\pagerange{\pageref{firstpage}--\pageref{lastpage}}
\maketitle

%%%%%%%%%%%%%%%%%%%%%%%%%%%%%%%%%%%%%%%%%%%%%%%%%%%%%%%%%%%%%%
\begin{abstract}
Mapping the small-scale structure of the universe through gravitational lensing is a promising tool for probing the particle nature of dark matter. Curved Arc Basis (CAB) has been proposed as a local lensing formalism in galaxy clusters, with the potential to detect low-mass dark matter substructure. In this work, we analyze the cluster lens Abell S1063 in search of dark matter substructure with the CAB formalism, using multi-band imaging data from JWST. We use two different source modeling methods: shapelets and pixel-based source reconstruction based on Delaunay triangulation. We find that source modeling systematics from shapelets result in a disagreement between CAB parameters measured from different filters. Source modeling with Delaunay significantly alleviates this systematic, as seen in the improvement in agreement across filters. We also find that inadequate complexity in source modeling can result in convincing spurious detections of dark matter substructure from strong gravitational lenses, as seen by our $\Delta \text{BIC} > 20$ measurement of a $M \sim 10^{10}$ $M_{\odot}$ subhalo with shapelets, a spurious detection that is not reproduced with Delaunay source modeling. We demonstrate that multi-band analysis with different JWST filters is key for disentangling source and lens model systematics from dark matter substructure detections.
\end{abstract}


%%%%%%%%%%%%%%%%%%%%%%%%%%%%%%%%%%%%%%%%%%%%%%%%%%%%%%%%%%%%%%
\section{Introduction}

The standard model of cosmology ($\Lambda$CDM) assumes the existence of Cold Dark Matter (CDM), which dominates the matter content of the universe and is essential to explaining an array of cosmological observables spanning cosmic epochs. These include observations of the cosmic microwave background (\citealt{Planck2020}) and large-scale structure of the universe (\citealt{Springel2006},\citealt{Gil-marin2016}, \citealt{Abbott2018}, \citealt{Troxel2018}, \citealt{Abolfathi2018}, \citealt{DESI2024iii}). While it has been remarkably successful at large scales, the cold dark matter paradigm has yet to be tested on small scales. Simulations with CDM predict an abundance of low-mass structures called subhalos residing in dark matter halos around galaxies (\citealt{deLucia2004}, \citealt{Chan2015}). At the same time, many alternative dark matter theories can reproduce our large-scale observations but predict distinct observational signatures at sub-galactic scales. These theories either modify the predicted density profiles and population statistics of subhalos, as seen in warm or self-interacting dark matter models (\citealt{Bode2001}, \citealt{Tulin2018}), or predict entirely different forms of substructure, such as interference patterns arising in ultralight axion dark matter (\citealt{Schive2014}). To this end, probing the small-scale structure of the universe can be the key to understanding the physics of dark matter.

Strong gravitational lensing is sensitive to the line-of-sight mass distribution between the source and the observer (\citealt{Mao1998}), making it a direct probe of the dark matter substructure. As predicted by general relativity, light rays bend around massive objects, such as galaxies or galaxy clusters, leading to the appearance of multiple distorted images of a background source. The mass distribution of the lens is often parametrized by a smooth profile. A dark matter clump, either within the lens or along the line of sight, will cause a perturbation on top of this smooth lens model, allowing us to detect dark matter substructure gravitationally. To this day, a few subhalos have been detected or reported as candidate detections with this method in galaxy-galaxy strong lenses (\citealt{Vegetti2010}, \citealt{Hezaveh2016}, \citealt{Sengul2022_pert}, \citealt{Lange2024}).

Cluster lenses are even more powerful than galaxies, as they often host rich collections of multiply imaged, highly magnified, distorted arcs. However, studies of dark matter substructure in cluster lenses have been hindered by the computational cost of analyzing pixel-level data from their extensive fields of view. To this end, typical strong lens modeling of clusters has involved reducing the rich pixel-level data to only the positional information of lensed images (e.g., \citealt{Lagattuta2019}). However, recent work has made significant progress on incorporating pixel-level information in global cluster models (\citealt{Acebron2024}, \citealt{Broadhurst2024}). As a result, cluster substructure studies have been limited to statistical constraints on subhalo population abundances (\citealt{Natarajan2007}, \citealt{Natarajan2017}). While such studies give us important insights into the abundance of dark matter substructure, probing the density profiles of individual subhalos with direct detection techniques allows us to go a step further. By comparing the measured concentration, ellipticity, and effective density slope (\citealt{Sengul2022_epl}) of subhalos to results from simulations with different dark matter models, we can support or rule out dark matter particle candidates and place tighter constraints on the viable parameter spaces of models.

A promising approach for direct detection of cluster substructure is to employ a \textit{local} lensing formalism. Curved Arc Basis (CAB) (\citealt{Birrer2021}) was proposed as a local description of gravitational distortions based on the eigenvalues and eigenvectors of the local lensing Jacobian, \rtext{the curvature of the tangential eigenvector, and an orientation of the curved arc}. \cite{Sengul2023} showed with simulations that dark matter substructure along the line of sight of highly magnified lensed arcs in galaxy clusters can be detected and studied with the CAB formalism. They also had the first application of CAB to real data by modeling lensed images of the first cluster lens SMACS 0723 observed by the James Webb Space Telescope (JWST).

In this work, we exploit the power of multi-filter analysis of JWST data to study the source, lens, and foreground systematics involved in substructure searches with the CAB formalism. We describe the observations of Abell S1063 in Section \ref{sec:abell_s1063}. We detail our source and lens modeling methods, as well as substructure search in Section \ref{sec:methods}. We present our results of modeling a triply-lensed system in Abell S1063 with the CAB formalism and our substructure search in Section \ref{sec:results}. We present our concluding thoughts in Section \ref{sec:conclusions}.

%%%%%%%%%%%%%%%%%%%%%%%%%%%%%%%%%%%%%%%%%%%%%%%%%%%%%%%%%%%%%%
\section{Abell S1063}
\label{sec:abell_s1063}

Abell S1063 (also known as RXC-J2248-ID3) is a rich cluster lens at redshift $z_{\text{lens}} = 0.348$. It was first identified by \cite{Abell1989} and was later observed as part of the Cluster Lensing And Supernova survey with Hubble (CLASH) program (\citealt{Postman2012}). \cite{Caminha2016} used the latter imaging data together with spectroscopic follow-up observations with the VIsible Multi-Object Spectrograph (VIMOS) and Multi Unit Spectroscopic Explorer (MUSE) on the Very Large Telescope (VLT) to identify 47 lensed images and construct a global mass model. Most recently, \cite{Beauchesne2024} combined these observations with X-ray emission from \textit{Chandra} and velocity dispersion measurements of the member galaxies from MUSE to present an improved mass model.

%%%%%%%%%%%%%%%%%%%%%%%%%%%%%%%%%%%%%%%%%%%%%%%%%%%%%%%%%%%%%%
\subsection{Triply-imaged system 4}
\label{sec:system4}

We use the detailed studies of Abell S1063 to carefully select a set of multiple images best suited for study with the CAB formalism. We choose a set of three images identified as \textit{system 4} by \cite{Caminha2016} for a number of reasons:
\begin{itemize}
    \item The images are clearly separated, indicating that they do not cross critical curves, where CAB formalism breaks down. 
    \item A set of \textit{three} images mitigates lens modeling degeneracies.
    \item The common source is spectroscopically confirmed and measured to be at redshift $z_{\text{source}}=1.398$ (\citealt{Balestra2013}, \citealt{Richard2014}, \citealt{Caminha2016}).
    \item The images are highly magnified, with a high signal-to-noise ratio, which is important for a detailed substructure search.
    \item High-resolution JWST observations are available, and the images are visible across different filters. This allows for independent inference of the lens-mass distribution that should agree across filters. 
    \item The source has a complex, asymmetric morphology, making it more sensitive to perturbations in lensing due to substructure.
    \item The local field around each image is free of line-of-sight foregrounds and nearby cluster member galaxies that often contaminate the lensed light of multiple images and can bias CAB parameters, complicating perturber searches.
\end{itemize}

\begin{figure}
	\includegraphics[width=\columnwidth]{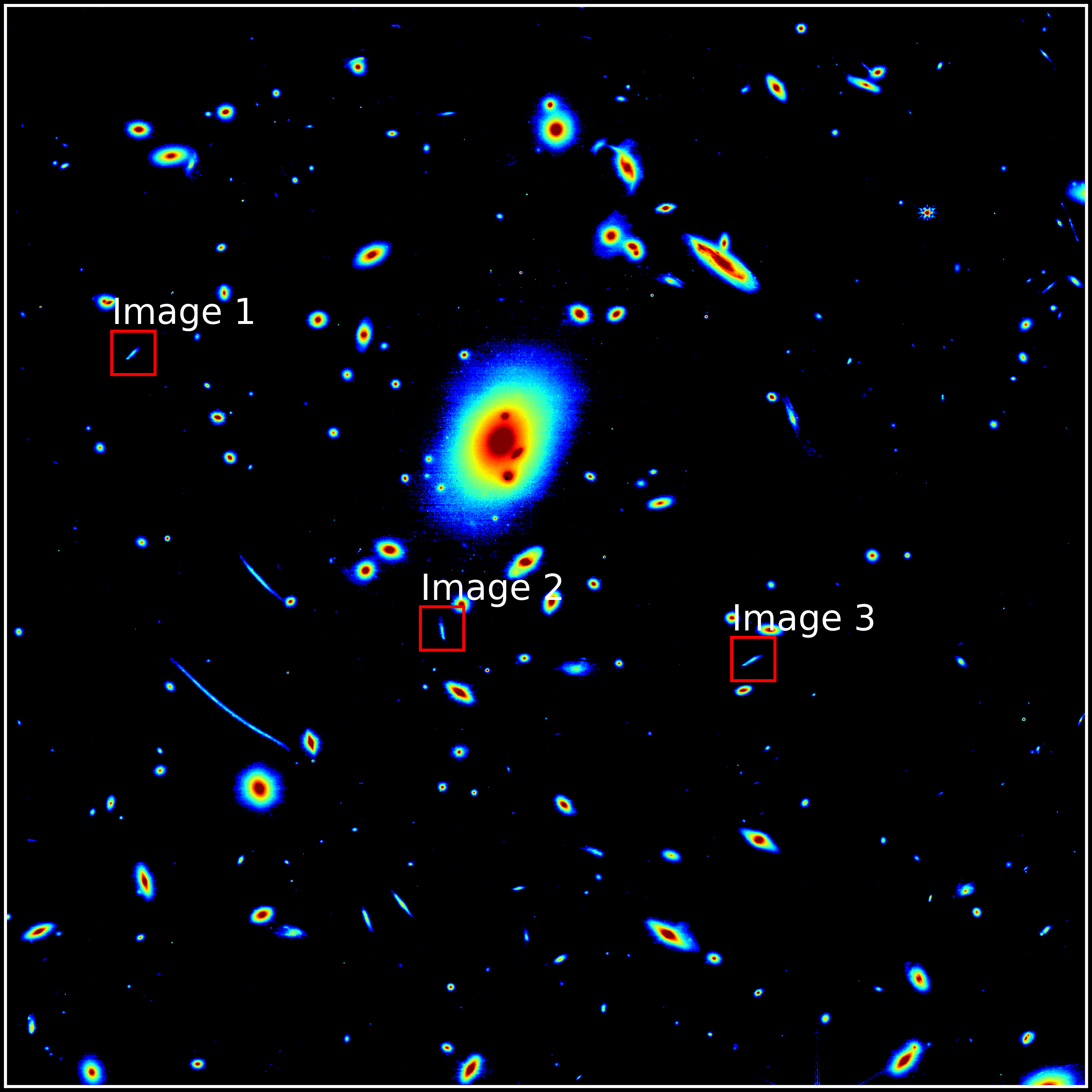}
    \caption{A 90 $\times$ 90 square arcsec field of view of galaxy cluster Abell S1063. Three lensed images of \textit{system 4} (see Section \ref{sec:system4}) are shown in red squares.}
    \label{fig:abell_s1063}
\end{figure}

The wider field of view of Abell S1063 is shown in Figure \ref{fig:abell_s1063}, where the positions of the three images of \textit{system 4} are indicated with red squares.

%%%%%%%%%%%%%%%%%%%%%%%%%%%%%%%%%%%%%%%%%%%%%%%%%%%%%%%%%%%%%%
\subsection{JWST Observations}
\label{sec:jwst_observations}

\begin{figure}
	\includegraphics[width=\columnwidth]{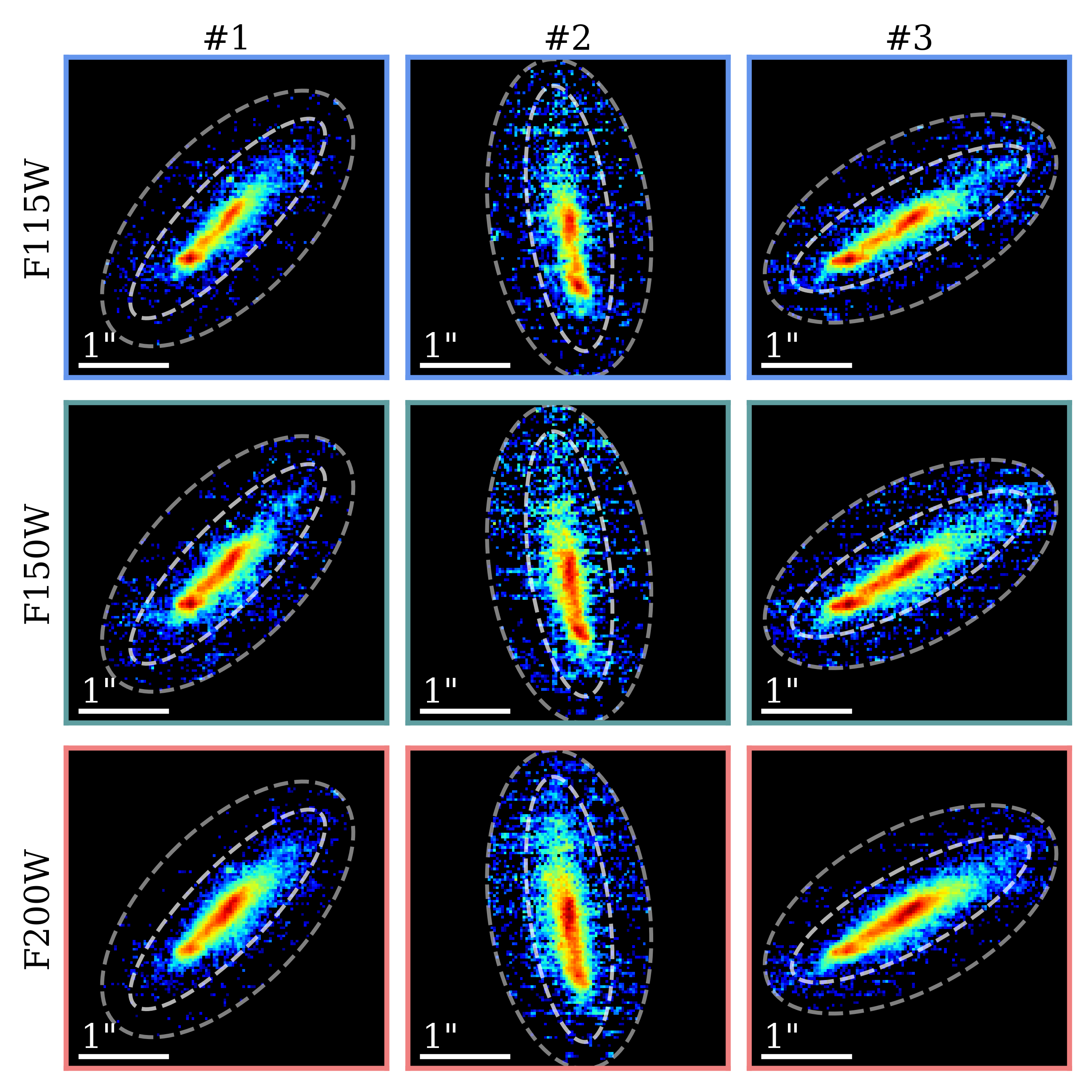}
    \caption{\rtext{The three lensed images of Abell S1063 \textit{system 4}, as seen in JWST NIRCam observations in filters F115W, F150W and F200W (\url{http://dx.doi.org/10.17909/mskr-6n50}).} The dashed ellipses show two different mask sizes applied to the image cutouts in our analysis, which we refer to as ``Small Masks" and ``Large Masks" in this work. \rtext{The color scale indicates $\log_{10}(\text{surface brightness)}$, with a fixed range of $[v_{\mathrm{max}}-1.4, v_{\mathrm{max}}]$ for each filter, where $v_{\mathrm{max}}$ is the maximum log-brightness across the three images in that filter}.}
    \label{fig:system4_all_images}
\end{figure}

\rtext{We use James Webb Space Telescope observations\footnote{\url{http://dx.doi.org/10.17909/mskr-6n50}} of Abell S1063 proposed in the General Observing Program under Program $\#1840$\footnote{\url{https://www.stsci.edu/jwst/science-execution/program-information?id=1840}} and taken on 24 September 2022 with the Near Infrared Camera (NIRCam).} The data is available in three short-wavelength and three long-wavelength filters. We only analyze the short-wavelength filters (F115W, F150W, F200W) as they offer a sharper point spread function and twofold pixel resolution compared to the longer-wavelength filters. Both features are important to our analysis, as the substructure search relies on distinguishing pixel-level changes in the brightness of lensed images. The broadband filters F115W, F150W, and F200W are sensitive to a range of non-overlapping frequencies centered at pivot wavelengths of 1.154, 1.501, and 1.988 $\mu$m, respectively. Thereby, each filter provides \rtext{correlated but varying} source morphologies lensed by the same mass distribution. Each filter's exposure times and signal-to-noise ratios are reported in Table \ref{tab:jwst_filters}. The image cutouts of \textit{system 4} in the three filters are shown in Figure \ref{fig:system4_all_images}. The data used in our analysis is publicly available on the Mikulski Archive for Space Telescopes (MAST)\footnote{\url{https://mast.stsci.edu/portal/Mashup/Clients/Mast/Portal.html}}.

\begin{table}
    \centering
    \label{tab:jwst_filters}
    \begin{tabular}{lccc}
        \hline
         & F115W & F150W & F200W \\
        \hline
        Pivot Wavelength [$\mu$m] & 1.154 & 1.501 & 1.988 \\
        Exposure Time [s] & 1,245 & 1,031 & 1,245 \\
        Signal-to-Noise Ratio & 110.1 & 130.9 & 140.6 \\
        Pixel Size [arcseconds] & 0.031 & 0.031 & 0.031 \\
        \hline
           \end{tabular}
        \caption{Observational properties of the JWST filters used in this work.}
\end{table}

We use the simulation tool \texttt{WebbPSF}\footnote{\url{https://www.stsci.edu/jwst/science-planning/proposal-planning-toolbox/psf-simulation-tool}} to model the point spread function (PSF) of each NIRCam filter. Typically, the PSF can also be calculated from pointlike sources in the field of view. However, we turn to simulation software because field stars are scarce along the line of sight of Abell S1063, located at a galactic latitude of $-60^{\circ}$, in a direction away from the galactic plane. 

%%%%%%%%%%%%%%%%%%%%%%%%%%%%%%%%%%%%%%%%%%%%%%%%%%%%%%%%%%%%%%
\section{Methods}
\label{sec:methods}

Strong lensing analysis requires simultaneous modeling of the source light and lens-mass distribution, obtained by minimizing residuals between the observed data and model reconstructions. Our data (see Figure \ref{fig:system4_all_images}) consist of elliptical cutouts from $120\times120$ pixel images of \textit{system 4} in Abell S1063 taken in three different filters. To model the cluster lensing, we use the local lensing formalism of CAB (\citealt{Birrer2021}), which has been shown to work well in describing highly magnified, extended arcs in cluster lenses, as well as to be sensitive to dark matter substructure (\citealt{Sengul2023}). To capture the light distribution of the background galaxy, we use two alternative source modeling methods: parametric shapelets (\citealt{Refregier2003}, \citealt{Birrer2015}, \citealt{Birrer2021_lenstronomy}) and pixel-based Delaunay triangulation (\citealt{Vegetti2009}, \citealt{Sengul2024}). To perform statistical fitting and search for dark matter substructure, we use Bayesian nested sampling with \texttt{dynesty} (\citealt{Speagle2020}).

%%%%%%%%%%%%%%%%%%%%%%%%%%%%%%%%%%%%%%%%%%%%%%%%%%%%%%%%%%%%%%
\subsection{Lens Modeling: CAB}

%%%%%%%%%%%%%%%%%%%%%%%%%%%%%%%%%%%%%%%%%%%%%%%%%%%%%%%%%%%%%%
\subsubsection{Angular Deflection in CAB}

\begin{figure}
	\includegraphics[width=\columnwidth]{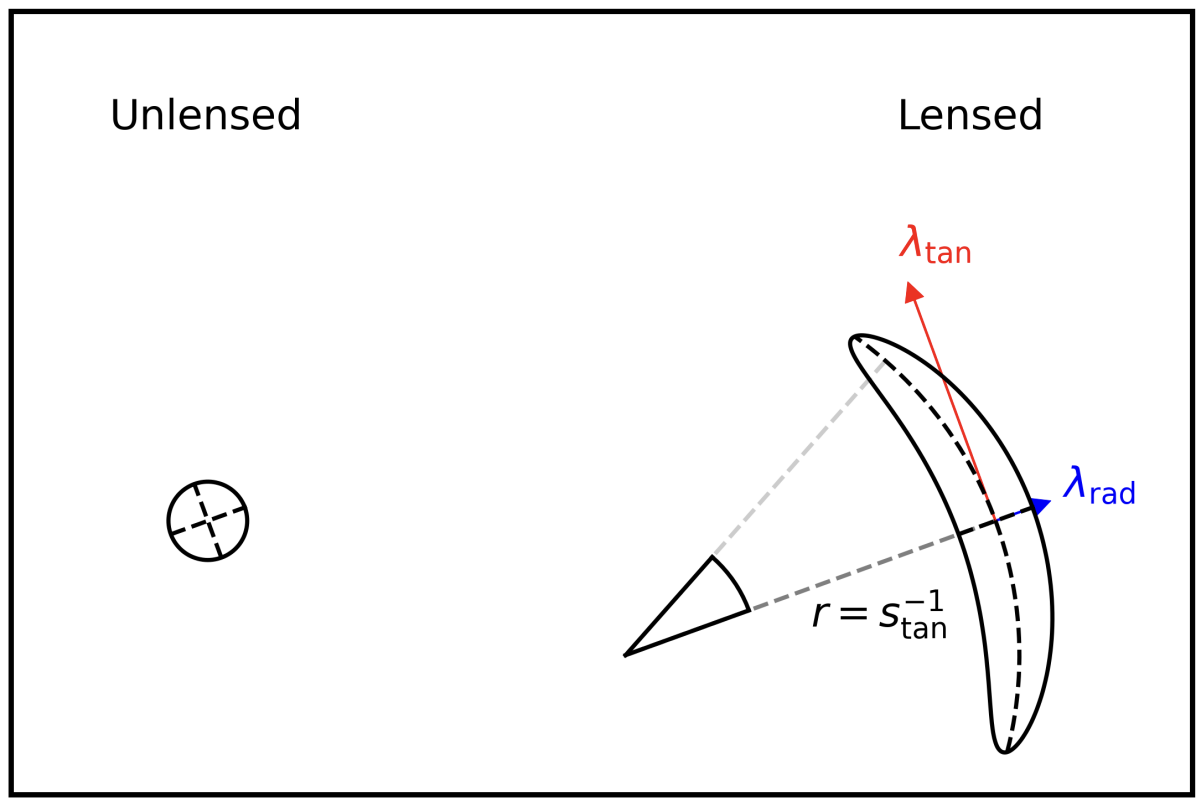}
    \caption{Illustration of CAB parameters (Figure from \protect\cite{Sengul2023}).}
    \label{fig:cab}
\end{figure}

The CAB formalism proposed by \cite{Birrer2021} uses the eigenvalues and eigenvectors of the local lensing Jacobian to describe distortions of extended arcs. This local model is fully equivalent to a global Singular Isothermal Sphere with a mass sheet transformation (MST)\footnote{
MST is a transformation of the form $1-\kappa' = \lambda (1-\kappa)$, where $\lambda \neq 0$ and $\kappa$ is the convergence field. Such transformations lead to a mass-sheet degeneracy as the shear signal is invariant under the addition of a constant surface mass density to the convergence field.} and can be described by four local parameters: tangential stretch $\lambda_\text{tan}$, radial stretch $\lambda_\text{rad}$, orientation $\phi$, and curvature $s_{\text{tan}}$. An illustration of the effect of a CAB distortion on a circularly symmetric source is shown in Figure \ref{fig:cab}.
The deflection angles of a CAB transformation centered at $\vec{\theta}_0$ can be expressed as:
\begin{align}
\label{eq:cab_alpha}
    \vec{\alpha}(\vec{\theta})=\lambda_{\mathrm{rad}}^{-1}\left[\vec{\alpha}_{\mathrm{SIS}}(\vec{\theta})-\vec{\alpha}_{\mathrm{SIS}}\left(\vec{\theta}_0\right)\right]+\left(1-\lambda_{\mathrm{rad}}^{-1}\right)\left(\vec{\theta}-\vec{\theta}_0\right),
\end{align}
where $\vec{\theta}_0$ gets mapped to the center position by construction, and the angular deflection $\alpha_{\text{SIS}}$ of a Singular Isothermal Sphere (SIS) centered at $\vec{\theta}_c$ is given by:
\begin{align}
\label{eq:sis_alpha}
    \vec{\alpha}_{\mathrm{SIS}}(\vec{\theta})=s_{\mathrm{tan}}^{-1}\left(1-\frac{\lambda_{\mathrm{rad}}}{\lambda_{\mathrm{tan}}}\right)\left(\frac{\vec{\theta}-\vec{\theta}_c}{|\vec{\theta}-\vec{\theta}_c|}\right).
\end{align}
The CAB formalism is especially relevant in cluster lenses, where the most massive component of the cluster dominates the strong gravitational lensing effect. \cite{Yang2020} have shown that, in some cases, lensed images in clusters may be well described by a constant convergence and shear, which is a special limit of the CAB as $s_\text{tan} \rightarrow \infty$. \citet{Sengul2023} first applied CAB to data, modeling JWST images of the cluster lens SMACS 0723, and presented a novel method for detecting and studying low-mass subhalos. In particular, with the local lensing formalism in hand, one can detect subhalos located near highly magnified arcs by discerning their perturbative corrections to the main lensing by the cluster. In mock JWST-quality simulations of the SMACS 0723 system, the authors showed that the CAB formalism could detect perturbers down to masses of $10^8 M_{\odot}$, allowing a measurement of their concentration, ellipticity, and redshift (\citealt{Sengul2023}).

%%%%%%%%%%%%%%%%%%%%%%%%%%%%%%%%%%%%%%%%%%%%%%%%%%%%%%%%%%%%%%
\subsubsection{Region of Validity}
\label{sec:region_of_validity}

CAB is a local formalism and is only able to perfectly capture the lensing distortions \rtext{for a global model that consists of a Singular Isothermal Sphere (SIS) and an allowed Mass Sheet Transform (MST),} as given in Equation (\ref{eq:cab_alpha}). Lenses that are more complex than \rtext{SIS and MST} will add corrections to CAB.

The ellipticity of the most massive cluster dark matter halo will induce derivatives along the tangential stretch of a curved arc. External shear will cause additional asymmetric distortions to the CAB lensing. These two effects can be accounted for by considering extensions of CAB incorporating tangential derivatives and shear, which add corrections to regular CAB away from the center of the image (\citealt{Birrer2021}). However, the true lensing in a cluster is more complex than CAB and its extensions: the real mass distribution of clusters does not precisely follow these parametric profiles, an effect which can become important locally; at the same time, nearby galaxies in the field of view of the lensed images can induce additional lensing distortions to the images. The masses and concentrations of these galaxies are often poorly constrained, and incorporating them into the lens modeling can make the problem computationally intractable. Therefore, it has been emphasized that CAB is a local formalism that should be used cautiously (\citealt{Birrer2021}). The region of local validity of a CAB approximation can easily be studied in simulations. However, the conclusions on the validity of CAB are heavily dependent on the specific configuration of our lensed images relative to the complex mass and light distribution of the cluster in question. In many cases, a detailed global model of a cluster lens may not be available. In this work, we take a general, empirical approach to illustrate the power of multi-band analysis with JWST in determining the region of CAB validity and disentangling lens model systematics. To this end, we perform our analysis in two different \textit{mask sizes} applied to our JWST images. Masks are elliptical boundaries on the lens plane within which all the pixels are included in our analysis and outside of which all the pixels are excluded, illustrated with white dashed elliptical contours in Figure \ref{fig:system4_all_images}. If the local lensing formalism is a good approximation, assuming all other systematics are negligible, the parameters measured in different filters should be consistent. If the formalism fails due to inadequacies of CAB away from image centers, we should expect to see a tension between the same parameters measured in different filters, since different filters will have different spatial extents of source light. However, systematics can also arise from source modeling. To distinguish the two sources of systematics, we employ two different source modeling methods, as discussed in Section \ref{sec:source_modeling}. 

%%%%%%%%%%%%%%%%%%%%%%%%%%%%%%%%%%%%%%%%%%%%%%%%%%%%%%%%%%%%%%
\subsection{Source Modeling}
\label{sec:source_modeling}

Source modeling is an important part of strong gravitational lensing analysis, but it is especially important for CAB studies. JWST cluster data often involves more strongly magnified and higher-resolution lensed images than typical galaxy-galaxy strong lensing systems, thereby demanding a more sophisticated source model. In this work, we explore the strengths and limitations of two different source modeling methods: parametric shapelets and pixel-based Delaunay triangulation.

%%%%%%%%%%%%%%%%%%%%%%%%%%%%%%%%%%%%%%%%%%%%%%%%%%%%%%%%%%%%%%
\subsubsection{Shapelets}

Shapelets (\citealt{Refregier2003}, \citealt{Birrer2015}, \citealt{Birrer2021_lenstronomy}) are an orthonormal set of weighted Hermite polynomials. A shapelet source model comprises a linear combination of the polynomial basis functions and is thereby fully specified by a list of shapelet coefficients, a reference scale, and a coordinate center. Shapelets are able to capture more complex light distributions than the common Sérsic (\citealt{Sersic1963}) profile of galaxies and have thus been commonmber of allowed basis functions $m$ through the relation $m = (n_{\text{max}}+1)(n_{\text{max}}+2)/2$. The size of the source is set by the reference scale parameter $\delta$. The smallest scale $l_{\text{min}}$ resolved by order $n_{\text{max}}$ and reference scale $\delta$ is given by the relation $l_{\text{min}} = \delta / \sqrt{n_{\text{max}}+1}$. Finally, the light peak position of the source model is controlled by the shapelet center $(x_0, y_0)$, which is a non-linear parameter in the model.

We determine the posterior distributions of $\delta$ and $(x_0,y_0)$ through nested sampling together with the lens model parameters, using uniform priors. At each sampling step, where the lens model is fixed, the light distribution on the lens plane is a simple linear combination of the shapelet basis function coefficients on the source plane. Thus, we can solve for the shapelet coefficients by a linear matrix inversion of an $N\times N$ matrix, where the length of the data vector $N$ is the total number of pixels in our image cutouts. As the basis order $n_{\text{max}}$ is a hyperparameter, we perform a Bayesian Information Criterion (BIC) analysis to find the optimal $n_{\text{max}}$ with the lowest $\text{BIC} = k \ln(N) + \chi^2$, where $k$ is the number of model parameters. This allows us to ensure that we use the optimal source complexity without overate BIC analyses for the shapelet basis order for each NIRCam filter, as the source model should naturally look different in different wavelength ranges in each filter, as well as for each lens model and mask size applied to image cutouts. All reported results use the shapelet $n_{\text{max}}$ that gives model fits with the lowest BIC score.

\rtext{In our fiducial analysis with shapelets, we use a Cartesian basis, which is commonly used in lens modeling but is better suited to spherical source morphologies. However, if the source is highly elliptical, the implicit prior set by our choice of basis can bias the results. For this reason, we repeat our analysis using source reconstruction with an elliptical shapelet basis, which introduces two additional non-linear parameters. We set uninformative priors on these two parameters, allowing source ellipticity $e$ and position-angle of the semi-major axis $\theta$ to vary between $e \in [0, 1]$ and $\theta \in [0, 2\pi]$.} \ptext{Henceforth, refer to these two shapelet source reconstruction schemes with different bases as ‘Cartesian shapelets’ and ‘elliptical shapelets’. When a statement applies to both methods, we refer to them collectively as `shapelets'.}

%%%%%%%%%%%%%%%%%%%%%%%%%%%%%%%%%%%%%%%%%%%%%%%%%%%%%%%%%%%%%%
\subsubsection{Delaunay Triangulation}
\label{sec:Delaunay_method}

We use a pixel-based source modeling method based on Delaunay triangulation as a non-parametric alternative to shapelets. The method was implemented in \cite{Sengul2024} and is similar to the adaptive-grid technique introduced by \cite{Vegetti2009} but differs in the algorithm for tesselation point selection. Most pixel-based modeling methods rely on various regularization methods to make pixel-level modeling more computationally tractable. However, regularization techniques can introduce new hyperparameters and modeling choices, such as demanding certain levels of smoothness in the source that may not always improve the accuracy of the source model.

In this work, we use the method described in \cite{Sengul2024} to carefully select tesselation pixels in a way that makes pixel-based modeling tractable without demanding source regularization. The algorithm relies on selecting more tesselation points in the regions of the image plane where the light morphology is more complex. We use the gradient of the local surface brightness as an indicator of source complexity. We calculate the normalized Hessian $|\text{det}\mathbf{H}_i|/p_i$ at each pixel, where $p_i$ is the pixel brightness value, and choose $N/4$ candidate tesselation points, where $N$ is the total number of unmasked pixels in our data. If neighboring pixels are chosen as tesselation points, we remove one of each pair of pixels to prevent degeneracies in the source model reconstruction. At each sampling step, where the lens model is fixed, the tesselation points $\{\mathbf{q}_i\}$ are mapped onto corresponding source plane points $\{\boldsymbol{w}_i\}$ through backward ray-tracing. The list of light amplitudes $\textbf{c} \equiv (c_1, c_2, ..., c_n)$ at these points are calculated through a non-negative least squares optimization, as negative source model amplitudes are not physical. Our optimization is implemented using the \texttt{scipy.optimize.nnls} package, which is based on the method developed by \cite{Bro1997}.

The list of vertex amplitudes fully determines our source model, as the source light can be reconstructed using Delaunay triangulation. On the source plane, we calculate the Delaunay triangulation of $\{\mathbf{w}_i\}$, dividing the plane into triangles. We calculate surface brightness values $I_s(\boldsymbol{w})$ within each triangle by interpolating between the values at the vertices of the triangles:
\begin{align}
    I_s(\boldsymbol{w}) =c_i D_{i j k}(\boldsymbol{w}) +c_j D_{k i j}(\boldsymbol{w}) +c_k D_{j k i}(\boldsymbol{w}),
\end{align}
where $(c_i, c_j, c_k)$ are the coefficients of the amplitudes at the three vertices $(\boldsymbol{w_i}, \boldsymbol{w_j}, \boldsymbol{w_k})$, and 
\begin{align}
    D_{i j k}(\boldsymbol{w}) \equiv \frac{\left(\boldsymbol{w}_1-\boldsymbol{w}_{k 1}\right)\left(\boldsymbol{w}_{j 2}-\boldsymbol{w}_{k 2}\right)-\left(\boldsymbol{w}_{j 1}-\boldsymbol{w}_{k 1}\right)\left(\boldsymbol{w}_2-\boldsymbol{w}_{k 2}\right)}{\left(\boldsymbol{w}_{i 1}-\boldsymbol{w}_{k 1}\right)\left(\boldsymbol{w}_{j 2}-\boldsymbol{w}_{k 2}\right)-\left(\boldsymbol{w}_{j 1}-\boldsymbol{w}_{k 1}\right)\left(\boldsymbol{w}_{i 2}-\boldsymbol{w}_{k 2}\right)}.
\end{align}
Any point $\boldsymbol{w}$ that does not fall within a Delaunay triangle is assigned the surface brightness of the nearest vertex. 

%%%%%%%%%%%%%%%%%%%%%%%%%%%%%%%%%%%%%%%%%%%%%%%%%%%%%%%%%%%%%%
\subsection{Bayesian Nested Sampling}

We determine the source and lens models that best fit our observed data using a Bayesian framework. We independently analyze the images taken in three different NIRCam filters described in Section \ref{sec:jwst_observations}. For each filter, our data vector $\mathbf{D}$ is an array of pixels in the image cutouts of \textit{System 4} shown in individual rows of Figure \ref{fig:system4_all_images}. Our model parameter vector $\mathbf{q}$ consists of either only lens or both lens and source parameters, depending on our source modeling method,
\begin{align}
    \mathbf{q} =
\begin{cases}
    (\mathbf{q}_{\text{lens}}, \mathbf{q}_{\text{source}}) & \text{if } S = S_{\text{Shapelets}}, \\
    \mathbf{q}_{\text{lens}} & \text{if } S = S_{\text{Delaunay}},
\end{cases}
\end{align}
since \ptext{Cartesian and elliptical shapelets involve three or five non-linear source model parameters, respectively}. Our model reconstruction $\mathbf{M}(\mathbf{q})$ is given by
\begin{align}
    \mathbf{M}(\mathbf{q}) = \mathcal{T}_{\text{PSF}}\{
    \mathcal{L}\left(\mathbf{q}_{\text {lens }}\right)[S\left(\mathbf{q}_{\text {source }}\right)] \},
\end{align}
where the lensing operator $\mathcal{L}\left(\mathbf{q}_{\text {lens }}\right)$ acts on the source light model $S\left(\mathbf{q}_{\text {source }}\right)$ to produce the lensed images, which are convolved with the PSF and then pixelized to an appropriate JWST-resolution grid, together denoted by $\mathcal{T}_{\text{PSF}}$ operator, to give realistic image reconstructions. The source model $S\left(\mathbf{q}_{\text {source }}\right)$ is found through either a shapelets or a Delaunay triangulation reconstruction detailed in Section \ref{sec:source_modeling}. The lensing operator $\mathcal{L}$ is fully determined by the angular deflections of the CAB model given by Equations (\ref{eq:cab_alpha}) and (\ref{eq:sis_alpha}). The point spread function operation $\mathcal{T}_{\text{PSF}}$ will depend on the NIRCam filter and determines the resolution of our observed lensed images in each filter. We implement all modeling in this work using \texttt{lenstronomy} (\citealt{Birrer2018}, \citealt{Birrer2021_lenstronomy}), a publicly available strong gravitational lensing package for modeling and simulations. 

The \textit{likelihood} of observing data $\mathbf{D}$ given a set of lens model parameters $\mathbf{q}$ is given by
\begin{align}
    P(\mathbf{D} \mid \mathbf{q})=\frac{\exp \left[-\frac{1}{2}(\mathbf{D}-\mathbf{M}(\mathbf{q}))^T \boldsymbol{\Sigma}_{\text {pixel }}^{-1}(\mathbf{D}-\mathbf{M}(\mathbf{q}))\right]}{\sqrt{(2 \pi)^{\operatorname{dim}(\mathbf{D})} \operatorname{det}\left(\boldsymbol{\Sigma}_{\text {pixel }}\right)}},
    \label{eq:likelihood}
\end{align}
where ${\Sigma}_{\text {pixel }}$ is the covariance matrix of the pixel errors, which combine the Poisson noise and instrumental read noise at each pixel in quadrature. The above expression assumes that the pixel errors are Gaussian. However, any systematics in the model $\mathbf{M}(\mathbf{q})$ can lead to deviations from a Gaussian distribution in the residuals ($\mathbf{D}-\mathbf{M}(\mathbf{q})$), potentially biasing posterior distributions of the parameters. Hence, a multi-filter analysis is crucial for illuminating any systematics and testing the robustness of our model.

The posterior distribution of model parameters is obtained by weighting prior distributions $P(\mathbf{q})$ by the likelihoods,
\begin{align}
    P(\mathbf{q} \mid \mathbf{D})=\int d \mathbf{q} \,  P(\mathbf{q}) \, P(\mathbf{D} \mid \mathbf{q}).
\end{align}
In this work, we always assume uniform priors on all model parameters, listed in Table \ref{table:priors}. We use \texttt{dynesty} (\citealt{Speagle2020}) to implement this Bayesian analysis framework.

\begin{table}
\centering
\renewcommand{\arraystretch}{1.2} % Adjust row height
\begin{tabular}{|p{2cm}|p{2cm}|p{2cm}|}
\hline
\textbf{Parameter Name} & \textbf{Units} & \textbf{Priors} \\ \hline
$\lambda_{\text{tan},1}$ & $-$ & [1.0, 7.0] \\ \hline
$s_{\text{tan},1}$ & arcsec$^{-1}$ & [$-0.5$, 0.5] \\ \hline
$\phi_1/\pi$ & $-$ & [0.0, 1.0] \\ \hline
$\lambda_{\text{rad},2}$ & $-$ & [0.0, 7] \\ \hline
$\lambda_{\text{tan},2}$ & $-$ & [$-7.0$, 0.0] \\ \hline
$s_{\text{tan},2}$ & arcsec$^{-1}$ & [$-1.0$, 1.0] \\ \hline
$\phi_2/\pi$ & $-$ & [$-0.5$, 0.5] \\ \hline
$\alpha_{2,x}$ & arcsec & [$-0.2$, 0.2] \\ \hline
$\alpha_{2,y}$ & arcsec & [$-0.2$, 0.2] \\ \hline
$\lambda_{\text{rad},3}$ & $-$ & [0.0, 7.0] \\ \hline
$\lambda_{\text{tan},3}$ & $-$ & [0.0, 7.0] \\ \hline
$s_{\text{tan},3}$ & arcsec$^{-1}$ & [$-1.0$, 1.0] \\ \hline
$\phi_3/\pi$ & $-$ & [$-0.5$, 0.5] \\ \hline
$\alpha_{3,x}$ & arcsec & [$-0.2$, 0.2] \\ \hline
$\alpha_{3,y}$ & arcsec & [$-0.2$, 0.2] \\ \hline
$\delta$ & arcsec & [0, 0.5] \\ \hline
$x_0$ & arcsec & [$-0.2$, 0.2] \\ \hline
$y_0$ & arcsec & [$-0.2$, 0.2] \\ \hline
\end{tabular}
\caption{Priors on the smooth CAB and non-linear source model parameters (when applicable).}
\label{table:priors}
\end{table}

%%%%%%%%%%%%%%%%%%%%%%%%%%%%%%%%%%%%%%%%%%%%%%%%%%%%%%%%%%%%%%
\subsection{Substructure Search}
\label{sec:sub_search_methods}

We search for dark matter substructure following the method proposed in \cite{Sengul2023}, by exploiting the proximity of potential subhalos to highly magnified regions of the lensed images. If a substructure is present along the line of sight of one of the images, it will cause a perturbation to the gravitational lensing sourced by the cluster, thereby referred to as a perturber. This signal would be picked up by a lens model with a dark matter subhalo on top of the CAB model. \cite{Sengul2023} showed in JWST-quality simulations that the CAB can be used to detect subhalos down to masses of $\sim 10^8 M_{\odot}$ and to measure their concentration, ellipticity, and redshift parameters.

We model the mass distribution of a dark matter substructure as a Navarro-Frenk-White profile (\citealt{NFW1996}, \citealt{NFW1997}), given by
\begin{align}
    \rho_{\mathrm{NFW}}(r)=\frac{\rho_{\mathrm{s}}}{\frac{r}{r_{\mathrm{s}}}\left(1+\frac{r}{r_{\mathrm{s}}}\right)^2},
\end{align}
which we parametrize in terms of $M_{200}$, the mass enclosed within the virial radius $R_{200}$, the radius at which the subhalo density is 200 times the critical density of the universe, and concentration $c$, defined by the ratio $c \equiv \frac{R_{200}}{r_\mathrm{s}}$. We treat the mass, concentration, and $(x_\mathrm{p}, y_\mathrm{p})$ position of the perturber as free parameters in our lens model. To this end, we are agnostic to the nature of the substructure and do not assume, for example, that the subhalo follows a CDM mass-concentration relation. We place uniform priors on each subhalo parameter listed in Table \ref{tab:perturber_priors} and search for a subhalo within $\pm 0.7"$ from the brightest image pixel. We constrain the subhalo search to this region since the lensing effect of the perturbers is most discernible in regions with large gradients in the light distribution.

\begin{table}
\centering
\begin{tabular}{|c|c|c|c|c|}
\hline
\textbf{Parameter} & \textbf{Initial} & \textbf{Wide} & \textbf{Narrow} & \textbf{Units} \\ \hline
$\mathrm{M}_{200}$        & $[0, 1\times10^{4}]$ & $[0, 1\times10^{4}]$ & $[0, 1\times10^{4}]$ & $\times 10^7$ $\mathrm{M}_{\odot}$ \\ \hline
$c$       & $[1, 100]$ & $[1, 1000]$ & $[1, 500]$ & $-$ \\ \hline
$x_{\mathrm{p}}$     & $[-0.7, 0.7]$ & $[-0.3, 0.3]$ & $[-0.1, 0.2]$ & arcsec ["] \\ \hline
$y_{\mathrm{p}}$     & $[-0.7, 0.7]$ & $[0.1, 0.8]$ & $[0.4, 0.75]$ & arcsec ["] \\ \hline
\end{tabular}
\caption{Priors used for subhalo parameters.}
\label{tab:perturber_priors}
\end{table}

\begin{figure}
	\includegraphics[width=\columnwidth]{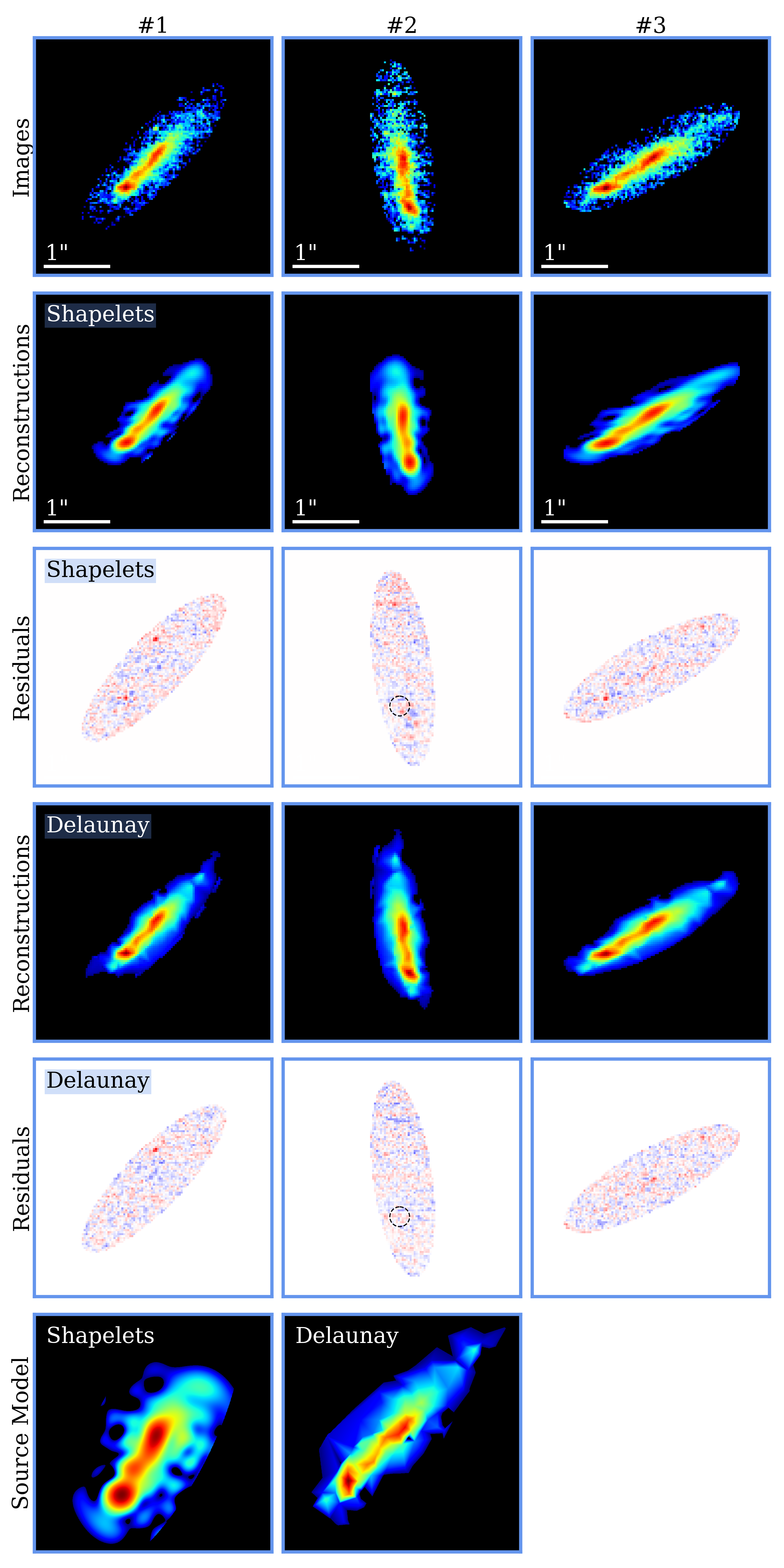}
    \caption{Model reconstructions of F115W Small Masks images with Delaunay and \ptext{Cartesian shapelet} source modeling methods. The shapelet model shown is with the optimal shapelet-order parameter $n_{\mathrm{max}} = 14$. The position of the spurious perturber detected in the substructure search with \ptext{Cartesian shapelet}s is shown with a black circle.}
    \label{fig:f115w_model}
\end{figure}

\begin{figure}
	\includegraphics[width=\columnwidth]{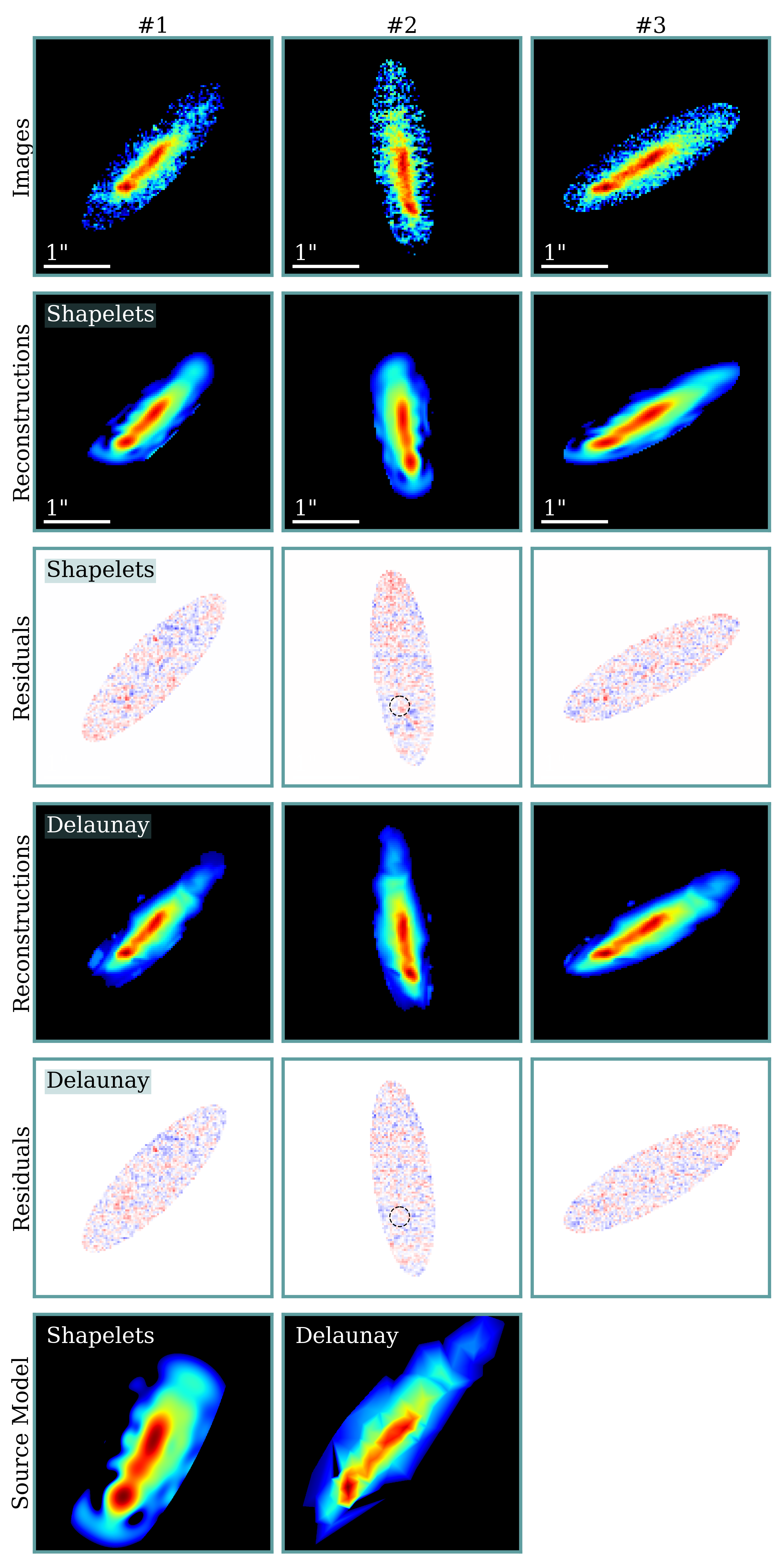}
    \caption{Model reconstructions of F150W Small Masks images with Delaunay and \ptext{Cartesian shapelet} source modeling methods. The shapelet model shown is with the optimal shapelet-order parameter $n_{\mathrm{max}} = 13$. The position of the spurious perturber detected in the substructure search with \ptext{Cartesian shapelet}s is shown with a black circle.}
    \label{fig:f150w_model}
\end{figure}

\begin{figure}
	\includegraphics[width=\columnwidth]{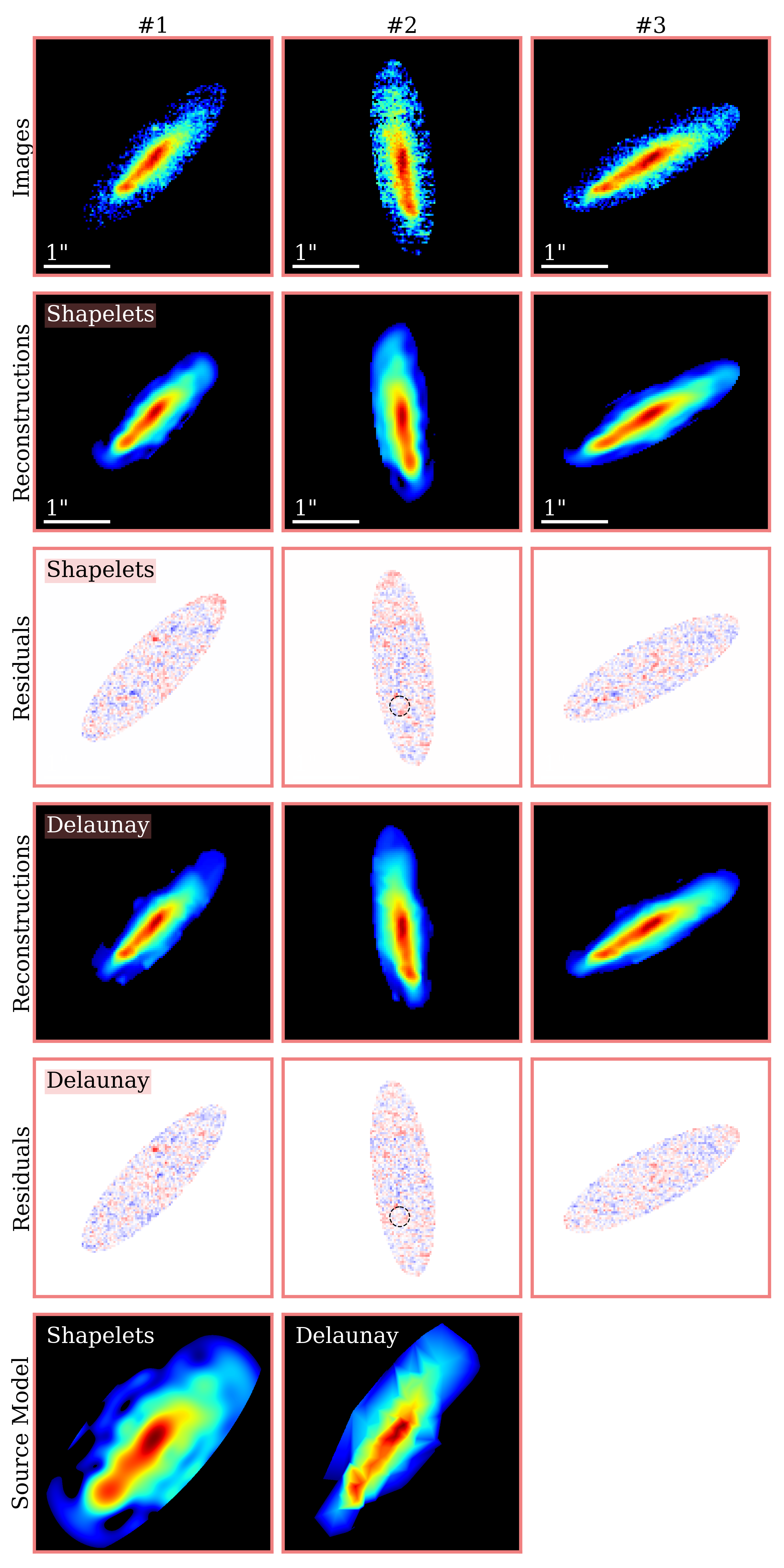}
    \caption{Model reconstructions of F200W Small Masks images with Delaunay and \ptext{Cartesian shapelet} source modeling methods. The shapelet model shown is with the optimal shapelet-order parameter $n_{\mathrm{max}} = 14$. The position of the spurious perturber detected in the substructure search with \ptext{Cartesian shapelet}s is shown with a black circle.
}
    \label{fig:f200w_model}
\end{figure}

%%%%%%%%%%%%%%%%%%%%%%%%%%%%%%%%%%%%%%%%%%%%%%%%%%%%%%%%%%%%%%
\subsubsection{Source Regularization}

Pixel-based source modeling methods commonly use regularization of the source light distribution (\citealt{Warr2003}, \citealt{Koop2005}, \citealt{Suyu2006}), where smoothness conditions are imposed on the source light to speed up the lens inversion. To investigate the effect of regularization on substructure detection, we extend our Delaunay triangulation algorithm to include gradient regularization introduced in \cite{Warr2003}. The modified merit function is given by:
\begin{align}
    G=\chi_{i m}^2+\lambda_{\mathrm{reg}} G_L,
\end{align}
where $G_L$ is the gradient regularization term and the $\lambda_{\mathrm{reg}}$ is the regularization strength parameter. Hence, the linear source inversion problem becomes:
\begin{align}
    \label{eq:source_inversion}
    \mathbf{S}=[\mathbf{M}+\lambda_{\mathrm{reg}} \mathbf{H}]^{-1} \mathbf{D},
\end{align}
where $\mathbf{H}$ is the regularization matrix. On a rectangular grid, typical gradient regularization penalizes differences in counts between neighboring pixels. In this work, we implement gradient regularization on a triangular grid by minimizing the differences between neighboring node values in a Delaunay mesh. The regularization term $G_L$ is a sum over all Delaunay triangles of the local gradient at the triangle, weighted by the areas of the triangles:
\begin{align}
  G_L
  \;=\;
  \sum_{T \in {\mathrm{\{triangles\} }}}
  \int_{T}
    \|\nabla I_s(\mathbf{\mathbf{w}})\|^2 
  \,dA = \mathbf{c}^{T} H \mathbf{c}
\end{align}

The regularization matrix $\mathbf{H}$ is the matrix of all pairwise gradient dot products of $D_{ijk}(\mathbf{w})$:
\begin{align}
    H_{p, q}=\sum_{p, q \in \triangle} \int_{\triangle} \nabla D_{p(\Delta)}(\mathbf{w}) \cdot \nabla D_{q(\Delta)}(\mathbf{w}) d A.
\end{align}

As described in Section \ref{sec:Delaunay_method}, we perform the regularized source inversion (Eq. \ref{eq:source_inversion}) at each sampling step using a non-negative least squares optimization. We fit for the regularization parameter $\lambda_{\mathrm{reg}}$ together with the lens model parameters in our Bayesian nested sampling framework.
%%%%%%%%%%%%%%%%%%%%%%%%%%%%%%%%%%%%%%%%%%%%%%%%%%%%%%%%%%%%%%

\begin{figure*}
	\centering
	\includegraphics[width=\textwidth]{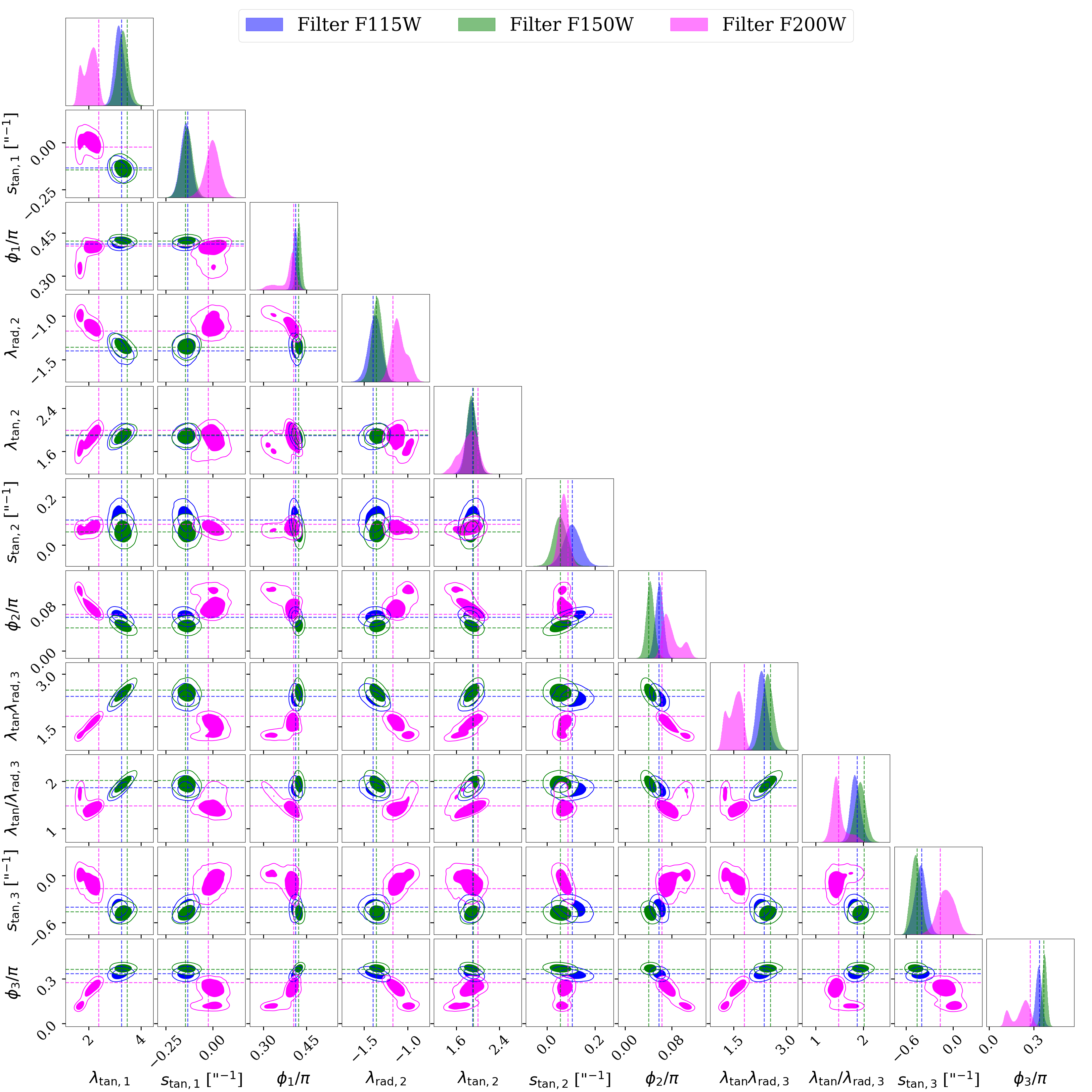}
    \caption{Posterior distribution of CAB parameters from \ptext{Cartesian shapelet} source modeling with $\times2$ boosted errors to account for modeling systematics. The shaded and unshaded 2D contours show $1\sigma$ and $2\sigma$ confidence intervals, respectively. The dashed lines show the best-fit parameters for each filter.}
    \label{fig:shapelets_boosted_noise}
\end{figure*}

\section{Results}
\label{sec:results}

\begin{figure*}
	\centering
	\includegraphics[width=\textwidth]{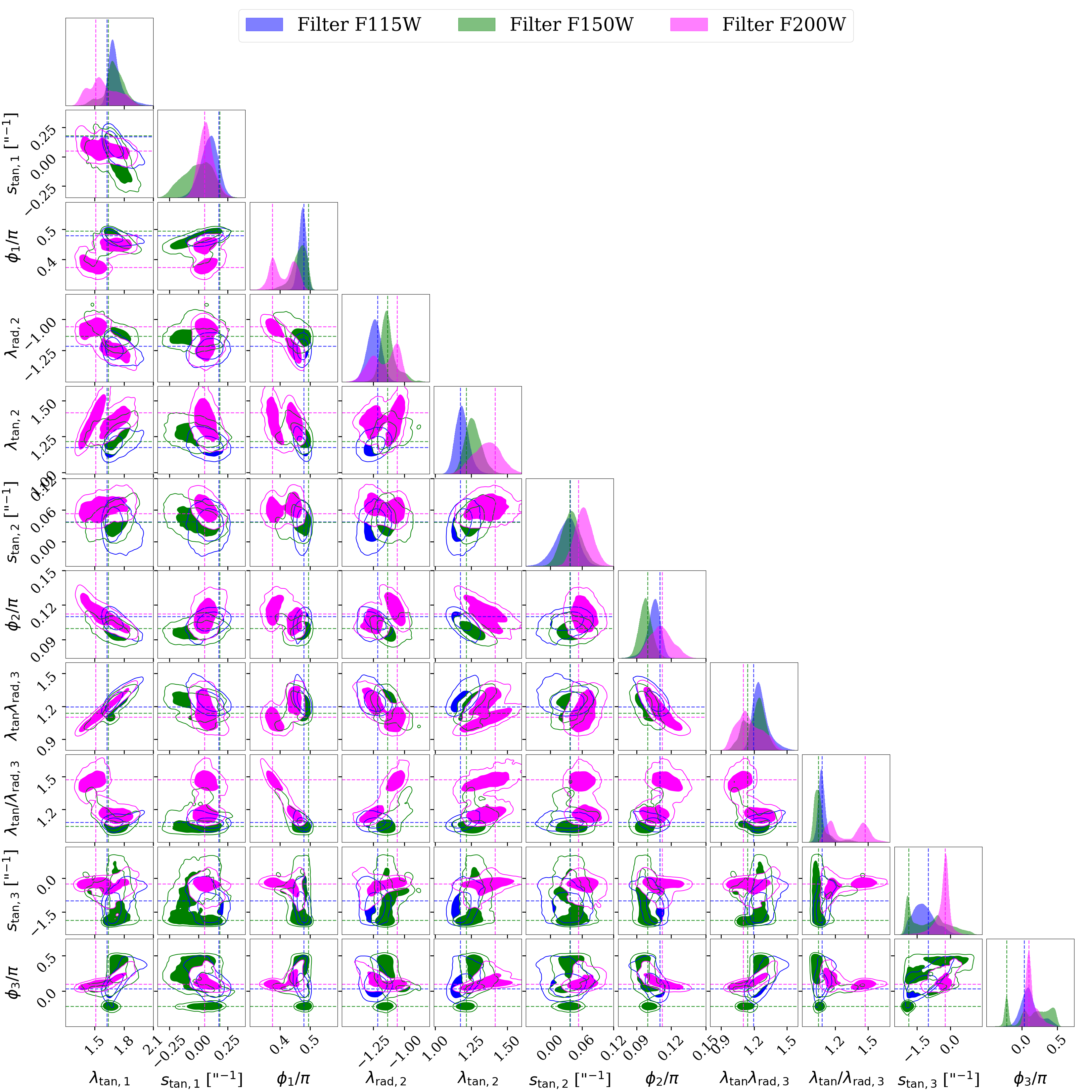}
    \caption{Posterior distribution of CAB parameters from Delaunay source modeling with $\times2$ boosted errors to account for modeling systematics. The shaded and unshaded 2D contours show $1\sigma$ and $2\sigma$ confidence intervals, respectively. The dashed lines show the best-fit parameters for each filter.
    }
    \label{fig:delaunay_boosted_noise}
\end{figure*}

\begin{figure*}
	\centering
	\includegraphics[width=\textwidth]{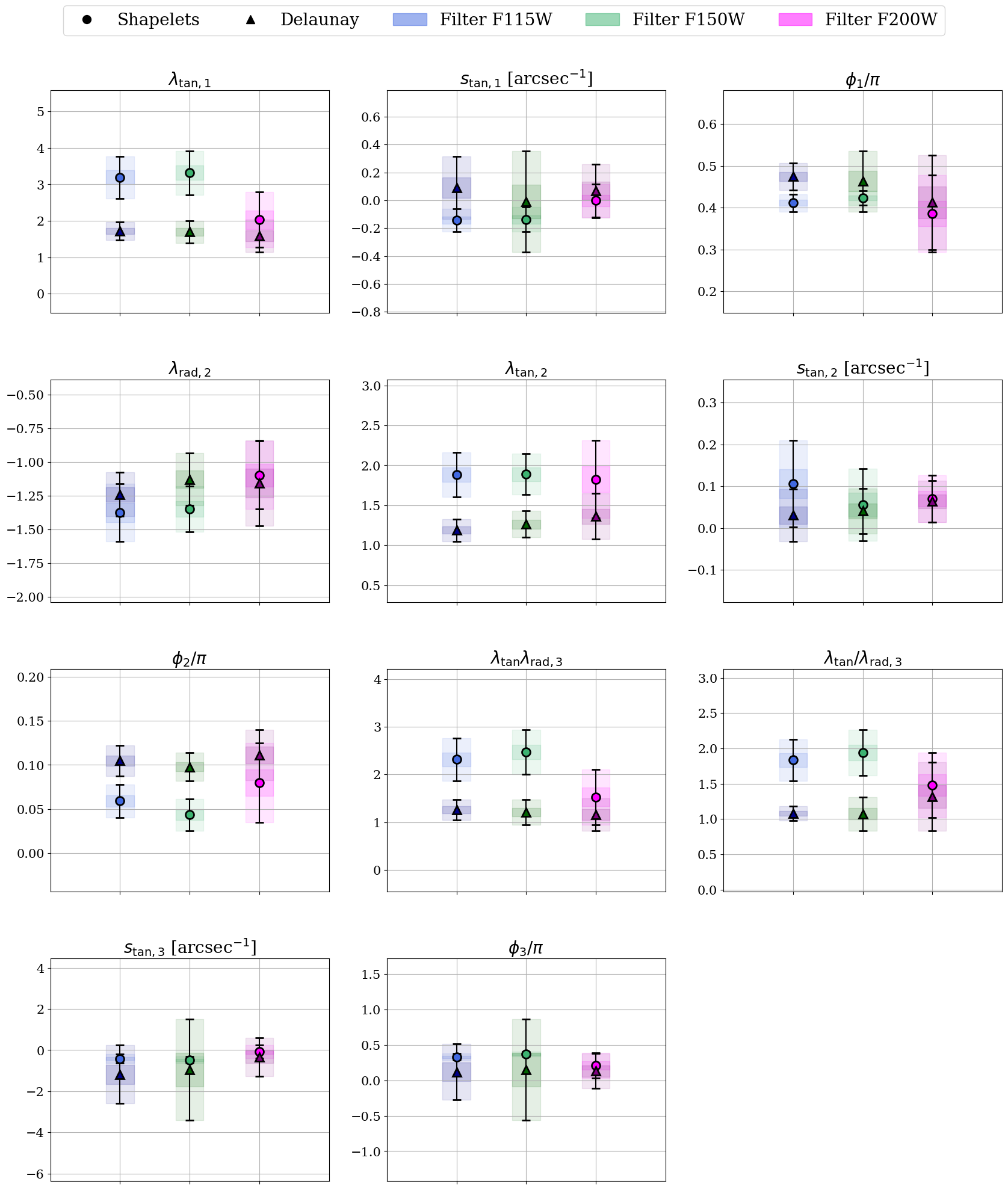}
    \caption{Means of CAB parameters obtained with \ptext{Cartesian shapelet} and Delaunay source modeling using Small Masks and $\times 2$ boosted errors. Darker and lighter shaded intervals indicate the $1\sigma$ and $3\sigma$ error bars, respectively.}
        \label{fig:mean_errors_boosted_noise}
\end{figure*}

In this section, we present our results of modeling \textit{system 4} in Abell S1063 with CAB and performing a dark matter substructure search. 

We first fit the lensed images with a smooth CAB model, using shapelet and Delaunay source modeling methods, and find improved agreement across filters with Delaunay, an indication of reduced systematics. Interestingly, our substructure search recovers a subhalo with shapelets but not with Delaunay, illustrating that insufficient source model complexity and inherent source regularization can lead to spurious substructure detections. We highlight the power of multi-band analysis in distinguishing dark matter substructure from source and lens model systematics.

%%%%%%%%%%%%%%%%%%%%%%%%%%%%%%%%%%%%%%%%%%%%%%%%%%%%%%%%%%%%%%
\subsection{Smooth CAB}

We show the best-fit model reconstructions and residuals of a smooth CAB model obtained with \ptext{Cartesian shapelet}s and Delaunay source modeling methods in Figures \ref{fig:f115w_model}, \ref{fig:f150w_model} and \ref{fig:f200w_model}, for filters F115W, F150W, and F200W, respectively. The \ptext{Cartesian shapelet} fits shown use the shapelet order parameters $n_{\mathrm{max}}$ optimized for each filter. Interestingly, the best-fit source models reconstructed using the two methods vary considerably. As perhaps expected from a pixel-based method, Delaunay can reproduce more detail in the complex source light distribution composed of two bright peaks connected with a thin twisted arc. The $\chi^2$ residuals are lower with Delaunay than with \ptext{Cartesian shapelet} source modeling, although it is difficult to fairly compare the two fits given the different numbers of degrees of freedom involved in the two source modeling methods.

%%%%%%%%%%%%%%%%%%%%%%%%%%%%%%%%%%%%%%%%%%%%%%%%%%%%%%%%%%%%%%
\subsubsection{Analysis with Large Masks}

We first analyze the $system$ $4$ images in three JWST filters using Large Masks, as shown in Figure \ref{fig:system4_all_images}. The posterior distributions of the CAB parameters obtained with \ptext{Cartesian shapelet} and Delaunay source modeling methods are shown in Appendix \ref{sec:large_masks_jwst_error}. Each JWST filter offers an independent realization of the PSF and a different source morphology, but probes the same line-of-sight mass distribution. Therefore, the CAB parameters should agree across different JWST filters, and any tension between lens parameters must arise from systematics in our source or lens modeling. To quantify this tension, we compute $ \mathrm{CT} = (\mu_i - \mu_j) / \sqrt{\sigma_i^2 + \sigma_j^2}$ for each parameter, where $\mu_i$ and $\sigma_i$ denote the mean and standard deviation of a parameter measured with the $i$th filter. Our results from both \ptext{Cartesian shapelet} and Delaunay source modeling methods show a significant $\mathrm{CT} \gtrsim 4$ average tension between lens parameters from different JWST filters. As discussed in Section \ref{sec:region_of_validity}, lens modeling systematics might arise from the failure of the local CAB approximation beyond its region of validity. This motivates us to shrink our mask sizes and consider smaller elliptical masks where we expect the CAB approximation to work better.

%%%%%%%%%%%%%%%%%%%%%%%%%%%%%%%%%%%%%%%%%%%%%%%%%%%%%%%%%%%%%%
\subsubsection{Analysis with Small Masks}

We repeat the multi-filter analysis with a smooth CAB model using Small Masks (see Figure \ref{fig:system4_all_images}). We show the posterior distributions of the CAB parameters obtained with \ptext{Cartesian shapelet} and Delaunay source modeling in Appendix \ref{sec:small_masks_jwst_error}. Reducing mask sizes leads to better agreement between filters with both source modeling methods, although a more significant improvement is seen with Delaunay. This shows that lens modeling systematics played an important role with Large Masks. Hence, multi-band analysis can be critical in determining the region of validity for a local lensing formalism.

Interestingly, Delaunay shows significantly better multi-filter agreement, with average tension between parameters at $\mathrm{CT} = 2.8$ as compared to $\mathrm{CT} = 3.9$ with \ptext{Cartesian shapelet}s. Increasing the number of tesselation points to $N/2$ (see Section \ref{sec:Delaunay_method}) in Delaunay source modeling further alleviates the tension to $\mathrm{CT} = 1.6$, a $\Delta \mathrm{CT} \gtrsim 2$ improvement from \ptext{Cartesian shapelet}s. However, the average CT scores still point to considerable tension between parameters. Thus, systematic errors dominate the pixel errors, violating the assumption that pixel errors are Gaussian in Equation (\ref{eq:likelihood}) and thereby biasing the posteriors. This motivates us to boost our JWST errros until the residuals from modeling failure become small compared to the boosted errors, burying systematics under Gaussian noise. 

%%%%%%%%%%%%%%%%%%%%%%%%%%%%%%%%%%%%%%%%%%%%%%%%%%%%%%%%%%%%%%
\subsubsection{Artificially Boosted Noise}
\label{sec:error_boost}

The tension between filters in our analysis with Small Masks indicates that our modeling systematics cannot fully handle the data quality of the JWST \textit{system 4} observations. To prevent systematics from biasing our measured CAB parameters, we boost the JWST error map by a factor of $2$ at each pixel. Such boosting of pixel errors was employed by \cite{Acebron2024} to limit the impact of imperfectly modeled areas on the lens model. We show the resulting posterior distributions from \ptext{Cartesian shapelet} and Delaunay source modeling in Figures \ref{fig:shapelets_boosted_noise} and \ref{fig:delaunay_boosted_noise}, respectively, and compare the means and errors of each parameter in Figure \ref{fig:mean_errors_boosted_noise}. We find that several parameters measured by \ptext{Cartesian shapelet}s are still in tension, with an average $\mathrm{CT} = 1.7$ between different JWST filters. In contrast, source modeling with Delaunay results in a $\sim 60\%$ reduction in the average CT score to $\mathrm{CT} = 0.74$, bringing all parameters within a $2 \sigma$ agreement between filters. We note that a two-fold boost in pixel errors leads to a factor of $\sim 3$ increase in the uncertainties of lens parameters. While error boosting does not solve the root cause of systematics, it allows us to adjust the data quality to a level that our systematics can successfully handle. With $\times 2$ boosted errors, Delaunay performs significantly better than \ptext{Cartesian shapelet}s, measuring consistent parameters. To reach the same level of consistency and thereby reliability with \ptext{Cartesian shapelet}s, one would need to further boost pixel errors, reducing the quality of the data and the amount of information we can extract from it.

%%%%%%%%%%%%%%%%%%%%%%%%%%%%%%%%%%%%%%%%%%%%%%%%%%%%%%%%%%%%%%

\subsection{Substructure search}

\begin{figure}
	\centering
	\includegraphics[width=\linewidth]{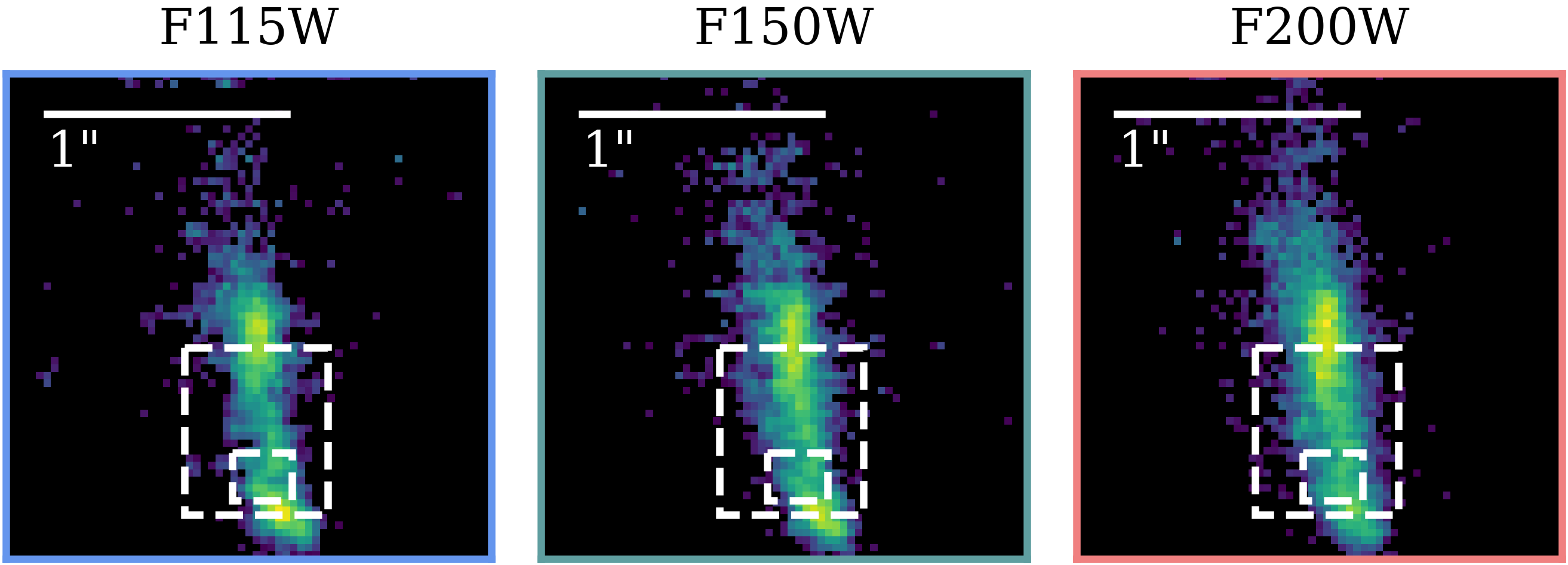}
    \caption{Positional priors used in the subhalo search in Image \#2. The full field of view represents the $1"\times1"$ region included in the preliminary substructure search. The smaller and larger dashed white rectangles show the positional prior bounds referred to as Narrow Prior and Wide Prior in this work.}
    \label{fig:perturber_pos_priors}
\end{figure}

We perform a substructure search using a smooth CAB model and both shapelet and Delaunay source modeling methods. We parametrize the dark matter substructure with an NFW profile and sample for its mass, concentration, and 2D position on the lens plane, assuming a lens redshift for the subhalo (see Section \ref{sec:sub_search_methods}). In our preliminary substructure search with Large Masks and \ptext{Cartesian shapelet} source modeling, we used wide uniform priors on all subhalo parameters, listed in Table \ref{tab:perturber_priors} as Initial Priors. We limited the initial subhalo positional priors to $x_i\in[-0.7, 0.7]$ arcseconds, where our coordinates are centered on the brightest image pixels, as proximity to bright regions is important for distinguishing the second-order lensing perturbation caused by subhalos. We searched for substructure in all three images of Abell S1063 \textit{system 4} and detected a subhalo in Image \#2 with \ptext{Cartesian shapelet}s. In our subsequent multi-filter analysis, we used tighter priors on the subhalo position centered at the preliminary detection location, referred to as Wide and Narrow Priors, listed in Table \ref{tab:perturber_priors} and illustrated in Figure \ref{fig:perturber_pos_priors}.

We find that source modeling systematics with shapelets can lead to convincing spurious detections of dark matter substructure. In particular, our substructure search with \ptext{Cartesian shapelet} source modeling results in a $\Delta\text{BIC}>20$ detection of a $\sim 10^{10}$ $\mathrm{M}_{\odot}$ subhalo located at consistent positions in all three filters. And yet, the recovered subhalo concentration differs significantly across the three filters, hinting that the subhalo detection may be spurious. Indeed, these results are not reproduced with Delaunay source modeling, which we find is more robust against spurious substructure detection arising from the inadequate complexity of the source model.

%%%%%%%%%%%%%%%%%%%%%%%%%%%%%%%%%%%%%%%%%%%%%%%%%%%%%%%%%%%%%%
\subsubsection{\ptext{Cartesian shapelet}s}
\label{sec:shapelets_substructure}
\begin{figure}
    \centering
    \includegraphics[width=\linewidth]{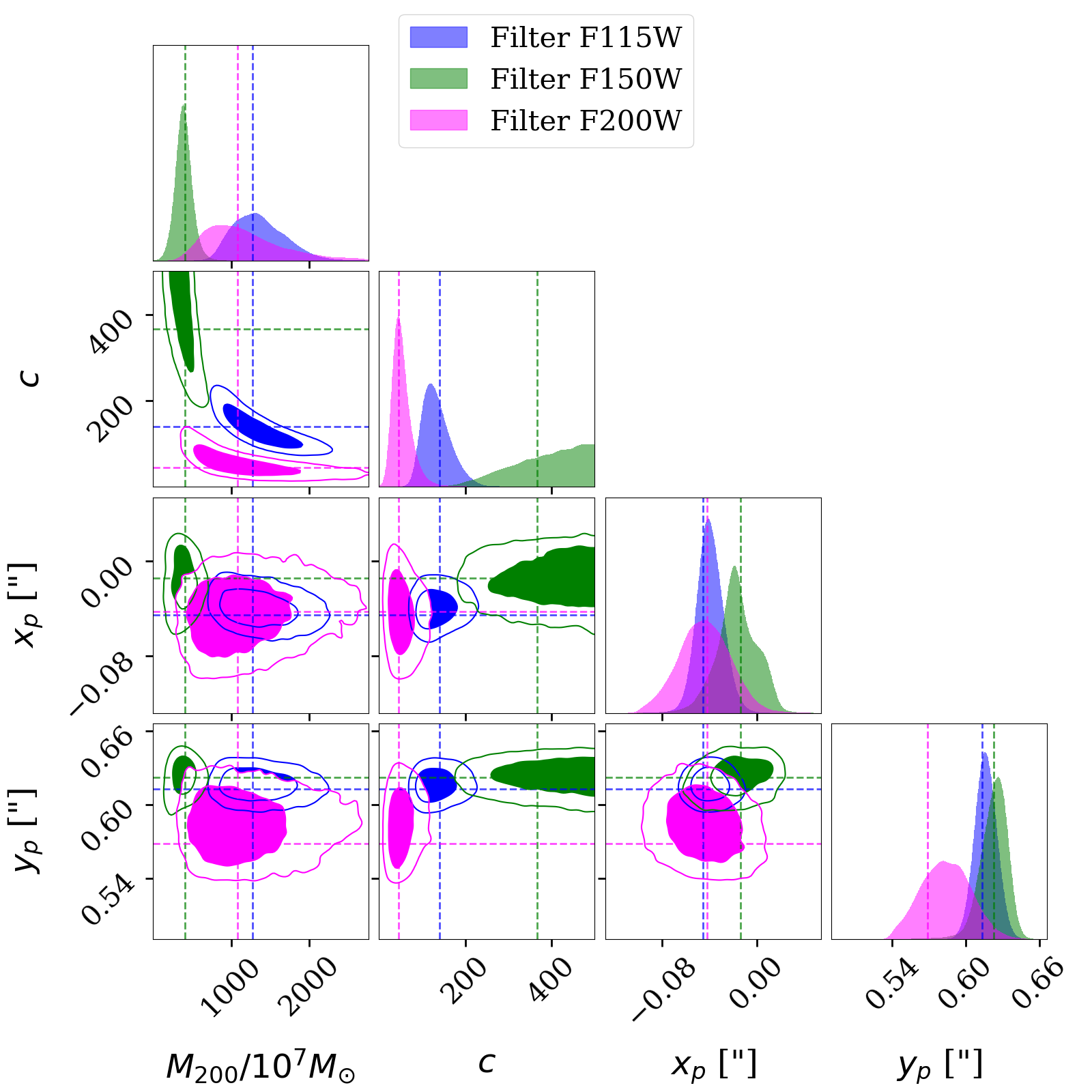}
    \caption{Posterior distribution of the NFW subhalo parameters recovered with \ptext{Cartesian shapelet}s using Large Masks with optimal shapelet-order parameters of $n_{\mathrm{max}}=$ 12, 14, and 15 for filters F115W, F150W, and F200W, respectively. The shaded and unshaded 2D contours show $1\sigma$ and $2\sigma$ confidence intervals, respectively. The dashed lines show the best-fit subhalo parameters for each filter.}
    \label{fig:nfw_overplot_large_masks}
\end{figure}

\begin{figure}
    \centering
    \includegraphics[width=\linewidth]{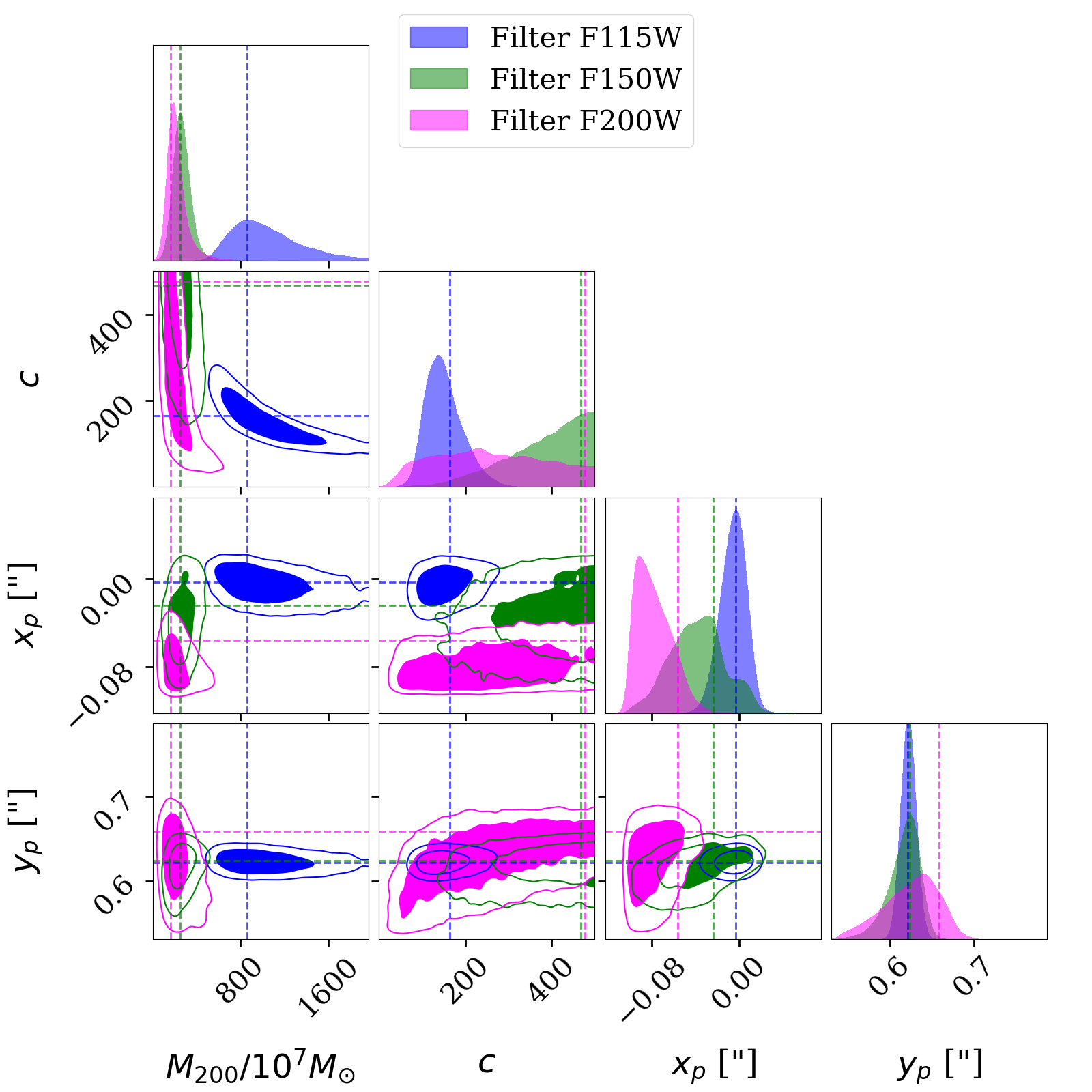}
    \caption{Posterior distribution of the NFW subhalo parameters recovered with \ptext{Cartesian shapelet}s using Small Masks with optimal shapelet-order parameters of $n_{\mathrm{max}}=$ 14, 13, and 14 for filters F115W, F150W, and F200W, respectively. The shaded and unshaded 2D contours show $1\sigma$ and $2\sigma$ confidence intervals, respectively. The dashed lines show the best-fit subhalo parameters for each filter.}
    \label{fig:nfw_overplot_small_masks}
\end{figure}

\begin{figure}
	\centering
	\includegraphics[width=\columnwidth]{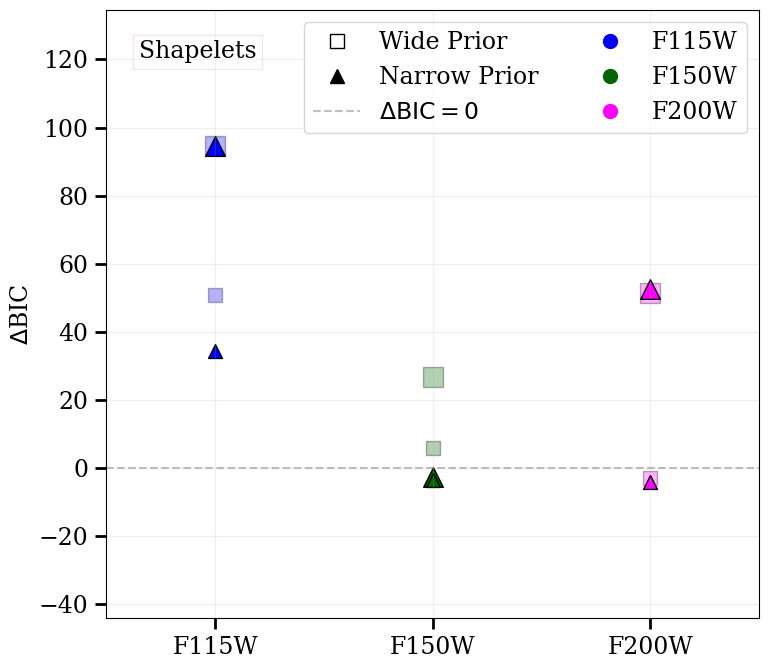}
    \caption{$\Delta$BIC values between CAB + NFW model fits compared to smooth CAB model using a \ptext{Cartesian shapelet} source model at optimal shapelet-order parameters of $n_{\mathrm{max}} = 14, 13, 14$ in Small Masks and $n_{\mathrm{max}} = 12, 14, 15$ in Large Masks for filter F115W, F150W, and F200W, respectively. Positive $\Delta$BIC indicates a preference for substructure over a smooth model. Smaller and larger marker sizes indicate fits to data with Small or Large Masks applied (see Figure \ref{fig:system4_all_images}), respectively. Fits with Wide or Narrow Priors on the perturber position (see Figure \ref{fig:perturber_pos_priors}) are shown with square or triangle markers, respectively. All three filters show strong $\Delta \mathrm{BIC} > 20$ preference for at least some substructure in Large Masks.} 
    \label{fig:shp_dBIC}
\end{figure}

\begin{figure}
	\centering
	\includegraphics[width=\columnwidth]{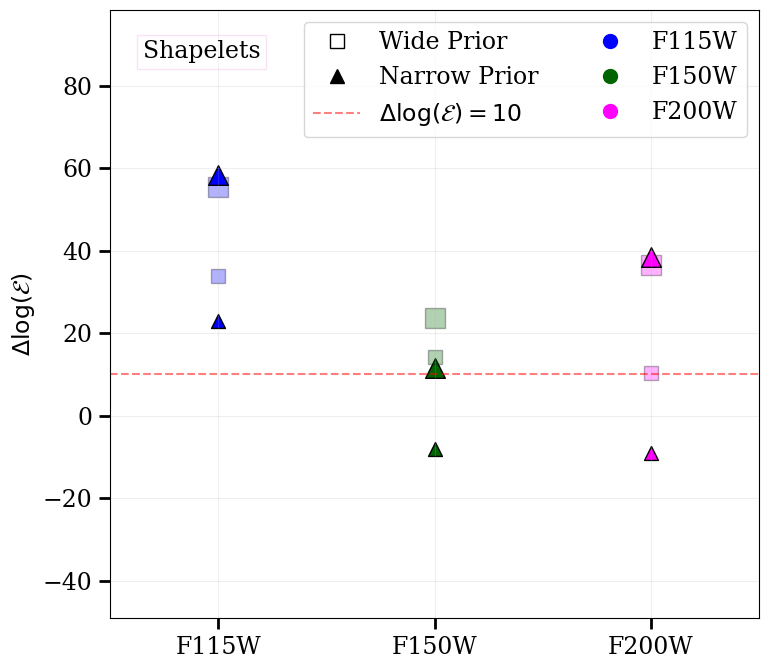}
    \caption{$\Delta \log(\mathcal{E})$ values between CAB + NFW model fits compared to smooth CAB model using a \ptext{Cartesian shapelet} source model at optimal shapelet-order parameters of $n_{\mathrm{max}} = 14, 13, 14$ in Small Masks and $n_{\mathrm{max}} = 12, 14, 15$ in Large Masks for filter F115W, F150W, and F200W. Symbol conventions are the same as in Figure \ref{fig:shp_dBIC}. All fits with Large Masks show strong statistical evidence $\log(\mathcal{E}) > 10$ for substructure.}
    \label{fig:shp_BF}
\end{figure}

We present the posterior distributions of the subhalo parameters recovered using \ptext{Cartesian shapelet} source modeling with Large Masks and Small Masks in Figures \ref{fig:nfw_overplot_large_masks} and \ref{fig:nfw_overplot_small_masks}, respectively. We find that all NFW subhalo parameters are consistent at least between two different JWST filters. However, the concentration parameter measured in F150W consistently hits an upper prior of $c=500$ and is in tension with the posteriors of the F115W and F200W filters.

As discussed in Section \ref{sec:source_modeling}, the complexity of the shapelet source model is controlled by the hyperparameter $n_{\text{max}}$ which should be optimized separately for the smooth and substructure lens models. We show the BIC curves of both models fitted to F115W, F150W, and F200W filters in Appendix \ref{sec:spurious_detection} and compare the performance of smooth and substructure models at the optimal $n_{\text{max}}$ of each model. The BIC statistic consistently favors the subhalo model over a smooth CAB model in both F115W and F200 filters. On the other hand, we don't see a significant difference in the BICs of smooth and substructure models for the F150W filter. It is possible to have a real substructure detection that is not reproduced in every filter if the SNR of those filters is insufficient to detect substructure (\citealt{Lange2024}). However, our lowest statistical significance for the subhalo is in the filter with an intermediate level of SNR (see Table \ref{tab:jwst_filters}), suggesting that it may be spurious.

We also show the posterior distribution of the subhalo parameters for a range of shapelet order parameters around the optimal $n_{\text{max}}$ in Appendix \ref{sec:spurious_detection}. The subhalo parameters are generally consistent across a broad range of shapelet order parameters. This suggests that even the most complex \ptext{Cartesian shapelet} source models may suffer from inadequate source model complexity or other systematics, leading to a preference for a substructure model. However, in both F115W and F200W filters, the $\Delta$BIC between smooth and perturber models tends to decrease with increasing $n_{\text{max}}$, another hint that the subhalo detection might be spurious.

We illustrate the detailed statistics of the subhalo detection with \ptext{Cartesian shapelet}s in terms of the $\Delta$BIC and the logarithm of the Bayes factor $\log(\mathcal{E})$ in Figures \ref{fig:shp_dBIC} and \ref{fig:shp_BF}, respectively. We repeated our substructure analysis using Small Masks and Large Masks and Wide and Narrow Priors imposed on the position of the subhalo in each of the three filters. With Large Masks and Narrow Priors, we consistently detect a subhalo with decisive statistics of $\Delta \mathrm{BIC} \gtrsim 20$ and $\log(\mathcal{E}) \gtrsim 20$. Shrinking the positional priors of the subhalo to the region enclosed by Narrow Priors leaves F115W and F200W results unaffected but reduces the statistical significance of the subhalo detection in F150W. This is because the positions recovered with Wide Priors in the intermediate filter disagree with the other two filters, and imposing the subhalo to be near the position detected by the other two filters reduces its statistical significance. Interestingly, switching from Large to Small Masks generally reduces both the $\Delta \mathrm{BIC}$ and $\log(\mathcal{E})$ of the subhalo detection in every filter. This suggests that lens modeling systematics arising from the failure of CAB formalism in larger masks also contribute to spurious subhalo detection.

Finally, we repeat our substructure search with $\times2$ boosted JWST errors (see Section \ref{sec:error_boost}), using Small Masks and Narrow Priors on the subhalo position. Notably, none of the filters continue to detect a subhalo with positive $\Delta \mathrm{BIC}$s. This provides further evidence that the detection of the subhalo was spurious. At the same time, the finding demonstrates the power of multi-band analysis. Looking for consistency in the lens parameters across filters allows us to decide whether we can trust our models. When tension across filters appear, we can use multi-band analysis to adjust the data quality to a level where systematics are drowned by boosted errors. To this end, we can prevent systematic effects from being mistaken for substructure detections.

%%%%%%%%%%%%%%%%%%%%%%%%%%%%%%%%%%%%%%%%%%%%%%%%%%%%%%%%%%%%%%

\begin{figure}
	\centering
	\includegraphics[width=\columnwidth]{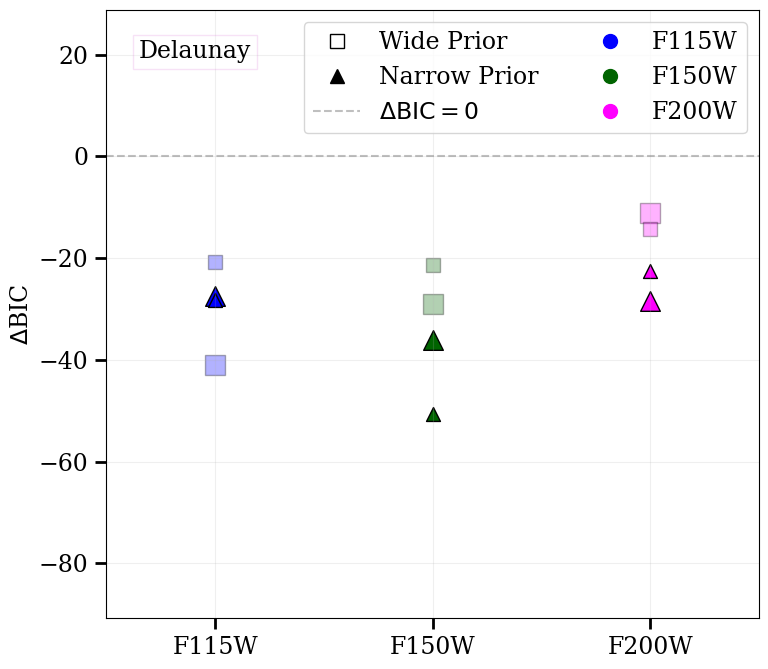}
    \caption{$\Delta$BIC values between CAB + NFW model fits compared to smooth CAB model using a Delaunay source model. Symbol conventions are the same as in Figure \ref{fig:shp_dBIC}. All combinations of fits have negative $\Delta \mathrm{BIC} < 0$, indicating a preference for a smooth model over a substructure.}
    \label{fig:dlny_dBIC}
\end{figure}

\begin{figure}
	\centering
	\includegraphics[width=\columnwidth]{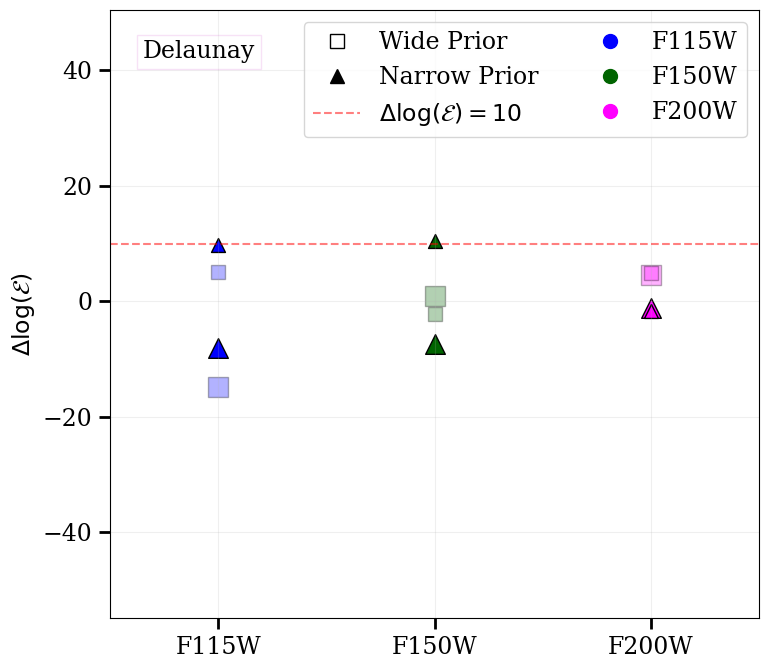}
    \caption{$\Delta \log(\mathcal{E})$ values between CAB + NFW model fits compared to smooth CAB model using a Delaunay source model. Symbol conventions are the same as in Figure \ref{fig:shp_dBIC}. None of the fits show strong evidence for substructure, as $\log(\mathcal{E}) < 10$ in all filters and mask sizes.}
    \label{fig:dlny_BF}
\end{figure}

\subsubsection{Delaunay}

We show the results of our substructure analysis relying on Delaunay source modeling in Figures \ref{fig:dlny_dBIC} and \ref{fig:dlny_BF}. Interestingly, we do not detect a subhalo in any of the three filters, as $\Delta \mathrm{BIC} < 0$ and $\log(\mathcal{E}) < 10$ consistently favor a smooth model over added substructure in both Large and Small Masks and when using either Narrow or Wide Priors on the subhalo position. This is consistent with our null detection of a subhalo using \ptext{Cartesian shapelet}s with $\times 2$ boosted errors. As a result, we see that Delaunay is more robust against spurious subhalo detections. This can be explained by the fact that Delaunay source modeling suffers from lower systematics, as seen by better agreement across filters when using Delaunay instead of shapelet source modeling.

\subsubsection{Elliptical Shapelets}

\rtext{We repeat our analysis with shapelets using an elliptical basis for source reconstruction. We show the resulting posterior distributions and subhalo detection statistics in Appendix \ref{sec:elliptical_shapelets}. As seen with Cartesian shapelets and Delaunay source modeling, the average CT score in Small Masks ($\mathrm{CT}=3.9$) is lower than that of Large Masks ($\mathrm{CT}=7.8$), supporting our conclusion that a growing failure of the local lens model leads to higher systematics in increasing mask sizes. However, switching to an elliptical shapelet basis does not alleviate tension in CAB parameters across filters, as CT scores are comparable between Cartesian and elliptical shapelets. In both Large and Small Masks, a substructure model is favored over a smooth CAB model in only one filter (F200W), with $\Delta \mathrm{BIC} \sim 50$ and $\sim 20$, respectively. Interestingly, in most filters and masks, the statistical significance of the subhalo detection is weaker as compared to Cartesian shapelets. This suggests that the spurious subhalo is statistically favored partially because of the failure of Cartesian shapelets to capture highly elliptical sources, arising from their implicit regularization that favors more spherical source morphologies. However, Delaunay still outperforms elliptical shapelets in robustness against spurious subhalos and agreement in parameters across filters.}

%%%%%%%%%%%%%%%%%%%%%%%%%%%%%%%%%%%%%%%%%%%%%%%%%%%%%%%%%%%%%%
\subsection{Source Regularization}

Shapelet source modeling is inherently regularized by the forms of the shapelet basis functions, which implicitly constrain the types of light distributions we can capture. For example, shapelets perform poorly in capturing highly peaked light distributions, thereby imposing a level of smoothness in the source. The spurious subhalo detection with shapelets may be a result of this inherent source regularization, as a spurious subhalo may compensate for the limited source complexity imposed by regularization.

\begin{figure}
    \centering
    \includegraphics[width=\linewidth]{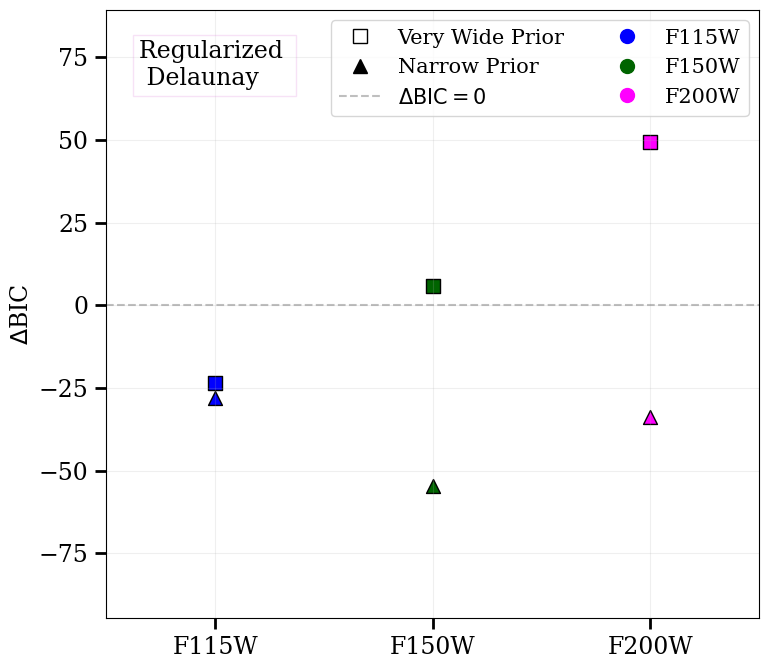}
    \caption{$\Delta$BIC values between CAB + NFW model fits compared to smooth CAB model using Small Masks and a Delaunay source model with gradient regularization (the parameter $\lambda_{\mathrm{reg}}$ is being varied in the range [0.1,1]). All Narrow Prior fits have negative $\Delta \mathrm{BIC} < 0$, indicating that regularized Delaunay also disfavors the subhalo detected with shapelets. However, a very wide perturber search recovers a subhalo with $\Delta \mathrm{BIC} > 0$ in two filters.}
    \label{fig:reg_dBIC}
\end{figure}

\begin{figure}
    \centering
    \includegraphics[width=\linewidth]{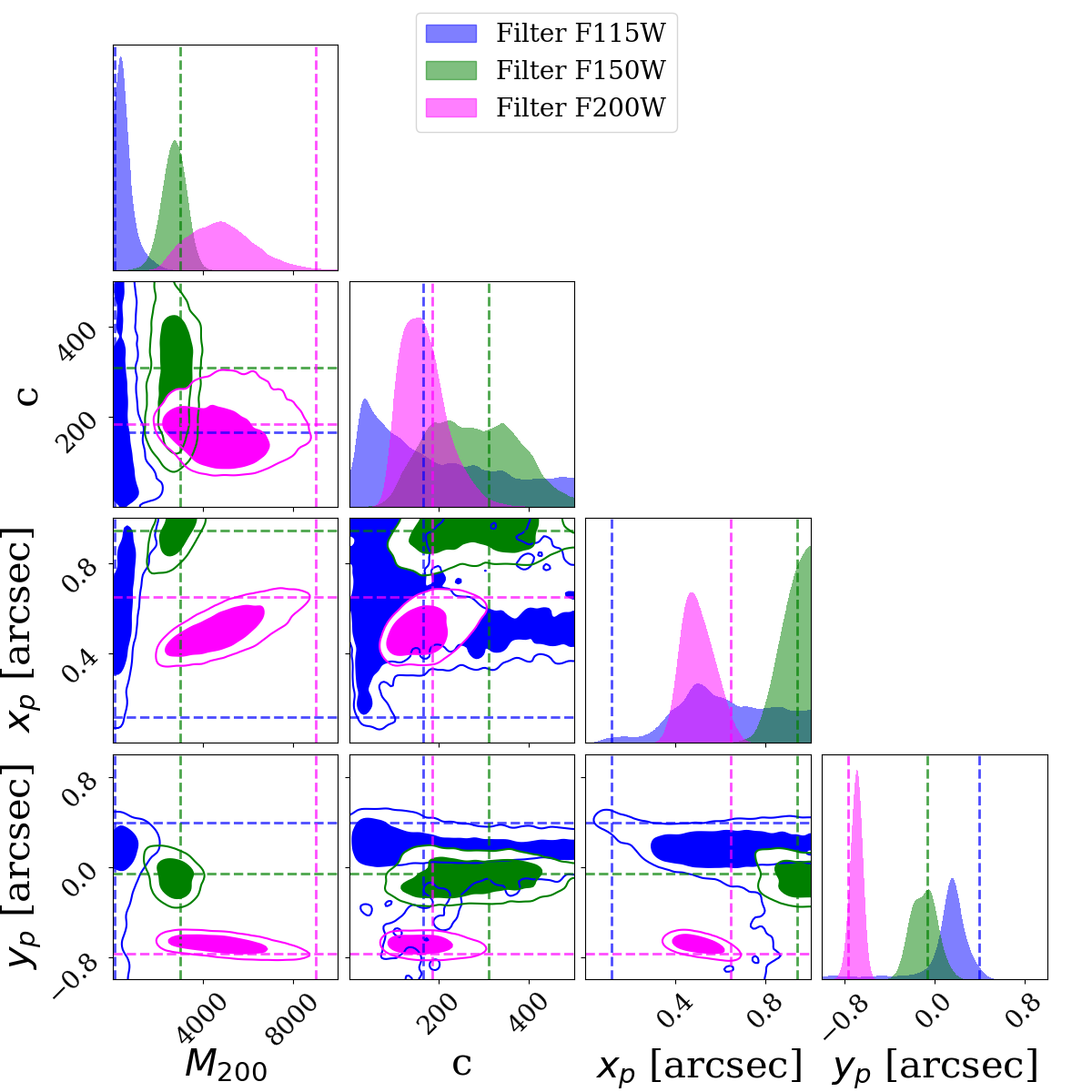}
    \caption{Posterior distribution of the NFW subhalo parameters recovered with regularized Delaunay using Small Masks and Very Wide Priors on the subhalo positions from filters F115W, F150W, and F200W. The shaded and unshaded 2D contours show $1\sigma$ and $2\sigma$ confidence intervals, respectively. The dashed lines show the best-fit subhalo parameters for each filter. We detect a subhalo in the F200W filter.}
    \label{fig:subhalo_reg}
\end{figure}

To test whether spurious subhalo detections can result from source regularization, we repeat our substructure search with regularized Delaunay source modeling, which we show in Figures \ref{fig:reg_dBIC} and \ref{fig:subhalo_reg}. To enforce some level of regularization in the source, we place linear priors of $\lambda_{\mathrm{reg}} \in [0.1, 1]$ on the regularization strength parameter. When searching for a subhalo with Narrow Priors on the position, we find that all substructure fits are disfavored by the BIC. Interestingly, however, a wide substructure search reveals a subhalo with $\Delta \mathrm{BIC} \sim 50$ in the F200W filter. We conclude that gradient regularization can indeed lead to spurious subhalo detections. \rtext{This finding is consistent with \cite{Nightingale2018} and \cite{Galan2024}, who demonstrate that constant source regularization, where the regularization strength is controlled by a single hyperparameter, can lead to poor reconstructions and biased parameters. They also show that such problems can be resolved with an adaptive regularization scheme (\citealt{Nightingale2018})}.

%%%%%%%%%%%%%%%%%%%%%%%%%%%%%%%%%%%%%%%%%%%%%%%%%%%%%%%%%%%%%%
\subsection{Comparison to Global Model}

\begin{figure*}
	\centering
	\includegraphics[width=0.9\textwidth]{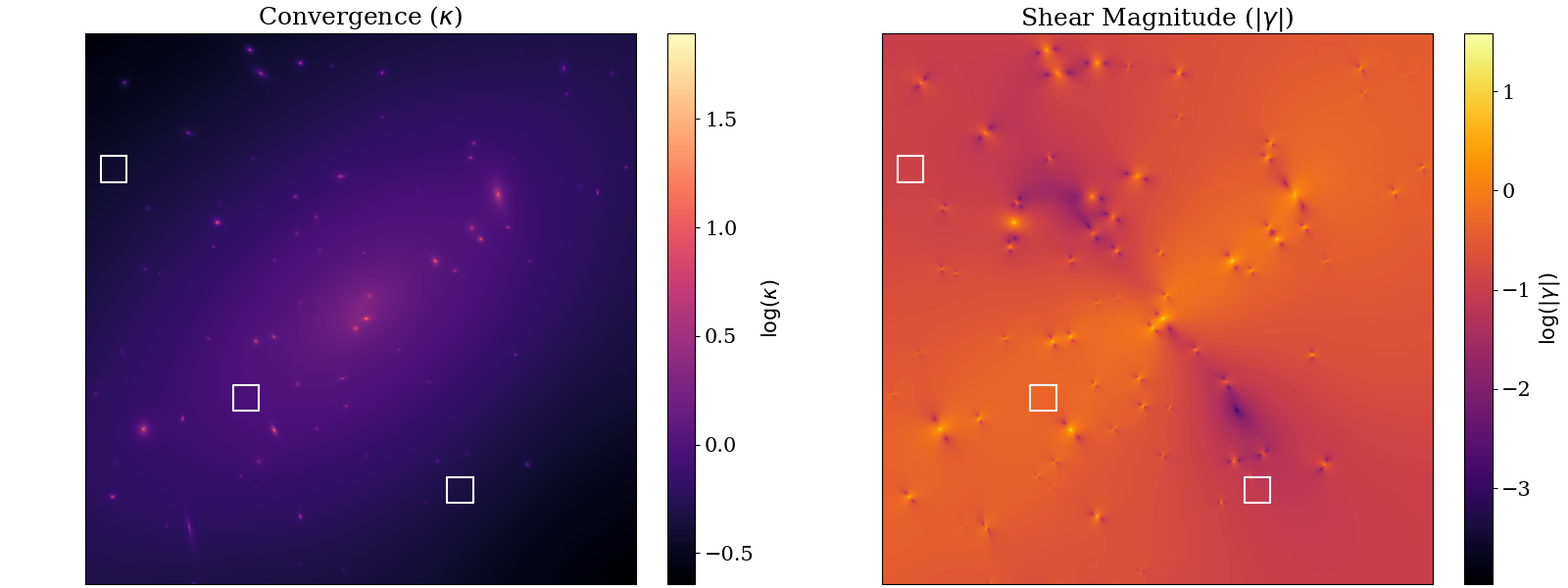}
    \caption{The convergence and shear maps from the best-fit global model of Abell S1063 presented in \protect\cite{Beauchesne2024}, which we use to derive the local CAB parameters predicted by the global model at the locations of the \textit{system 4} lensed images, indicated by the three white squares.}
    \label{fig:global_kappa_shear}
\end{figure*}

\begin{figure*}
	\centering
	\includegraphics[width=0.8\textwidth]{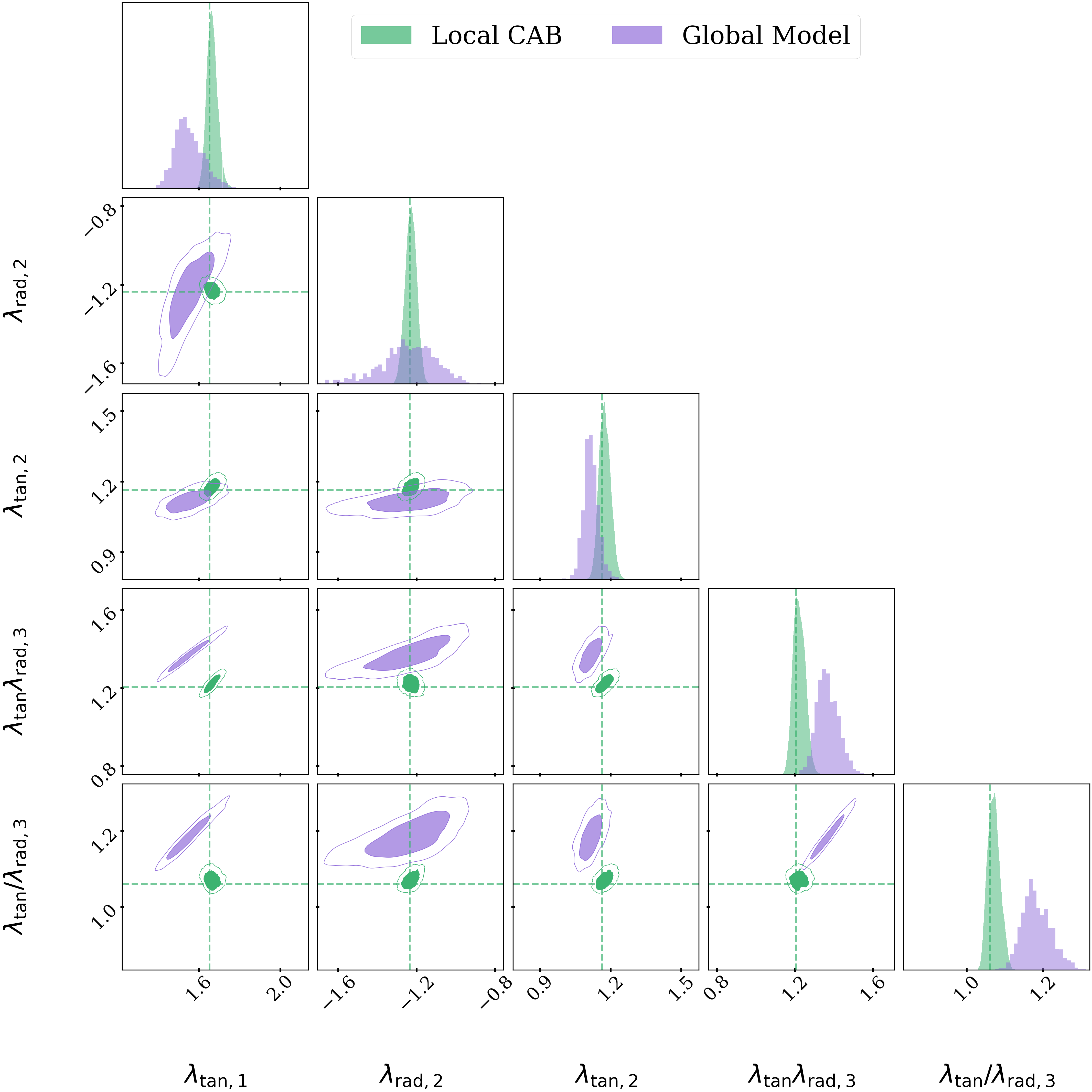}
    \caption{Comparison of CAB parameters $\lambda_{\mathrm{rad}}$ and $\lambda_{\mathrm{tan}}$ obtained by our local analysis and derived from the global model of Abell S1063 by \protect\cite{Beauchesne2024}. The shaded colored and line contours represent $68\%$ and $95\%$ confidence intervals. The local CAB results were obtained using Delaunay source modeling and Large Masks applied to filter F115W data. The global model contours were derived with Monte Carlo simulations, varying the parameters of the five most massive cluster lens components.}
    \label{fig:global_vs_local}
\end{figure*}
In this Section, we compare the CAB parameters obtained from our local analysis to the latest global cluster lens model of Abell S1063 by \cite{Beauchesne2024}. We show the best-fit global convergence and shear maps published by the authors in Figure \ref{fig:global_kappa_shear}, with the positions of the $\textit{system 4}$ images indicated with white squares. The cluster lens model is made up of 250 components, including two main dark matter halos near the brightest cluster galaxy (DM-BCG) and a high concentration of galaxies in the North-East (DM-NE), and three main gas components, all modeled as dual Pseudo-Isothermal Elliptical (dPIE) profiles (\citealt{Limousin2005}).

We present a comparison of our local CAB parameters with the CAB parameters predicted by the global model in Figure \ref{fig:global_vs_local}. \rtext{Due to a degeneracy to a mass sheet transform, only relative stretch parameters can be constrained by a local lensing analysis with CAB. Hence, we derive the stretch parameters from the global model relative to the first radial stretch, $\lambda_{\mathrm{r,1}}$, as constrained by our local analysis.} The green posteriors show the CAB stretches derived from our analysis of F115W data with Delaunay source modeling. The lilac contours show the $68\%$ and $95\%$ confidence intervals of the CAB parameters derived from the global model. The errors in the 28 sampled global model parameters of the five most massive components were propagated using Monte Carlo simulations of 1,000 random convergence and shear maps. For each map in the sample, we derived the CAB stretches by calculating the average eigenvalues of the lensing Jacobian in $1"\times1"$ cutouts centered at the \textit{system 4} images. We note that this procedure underestimates the errors in the CAB parameters derived from the global model, as it does not account for uncertainties in the velocity dispersion measurements of the individual cluster member galaxies.

We find good agreement between the global and local models in all but one of the CAB stretch parameters. Tension between the global and local models is expected, as the former does not incorporate pixel-level information and may be inaccurate at predicting local lensing distortions. At the same time, CAB parameters from local analysis are generally better constrained, even when underestimating the errors in CAB parameters from the global model. These results show that incorporating constraints from CAB in global lens modeling can improve the accuracy of state-of-the-art global cluster lens models. Recent work has made significant progress in incorporating pixel-level information in global cluster models, either by directly modeling bright lensed arcs (\citealt{Acebron2024}, \citealt{Broadhurst2024}) or by calculating higher-order moments of multiple images from pixel values without performing additional lens modeling (\citealt{Pignataro2021}, \citealt{Bergamini2021}, \citealt{Diego2022}, \citealt{Sharon2022}). Summary statistics from CAB analysis can complement these efforts, providing valuable information for global cluster models while maintaining the required computational efficiency to make the problem tractable.

%%%%%%%%%%%%%%%%%%%%%%%%%%%%%%%%%%%%%%%%%%%%%%%%%%%%%%%%%%%%%%
\section{Discussion \& Conclusions}
\label{sec:conclusions}

The CAB formalism is a promising method for probing dark matter substructure in strong cluster lenses. This work presents the first multi-filter study with CAB on real data to empirically test the method's robustness. We find that our results are highly dependent on the source modeling method. Shapelet source modeling leads to significant systematics that can bias the CAB parameters. In contrast, pixel-based source reconstruction with Delaunay triangulation is a more robust method that suffers from significantly less systematics, as seen in the better agreement of the CAB parameters across filters.

At the same time, CAB is a local formalism that should be used with caution. Overestimating the region of validity can lead to the failure of the CAB formalism, leading to lens modeling systematics that similarly bias the inferred parameters. A multi-band analysis in different mask sizes allows us to detect the failure of the local lensing formalism as a tension between lens parameters across filters. We demonstrate that CAB works better in smaller masks \rtext{for all source reconstruction methods}, as seen in reduced tension between parameters from different filters. Hence, we show the power of multi-band analysis as an empirical approach to determine the region of validity of this local formalism.

Source and lens modeling systematics have profound implications for substructure search. Our statistically significant detection of a subhalo with \rtext{Cartesian shapelets} tells a cautionary tale that source modeling systematics can lead to convincing substructure detections. \rtext{Switching to elliptical shapelets weakens the statistical preference for the spurious subhalo. This suggests that insufficient complexity of the source model can lead} to a preference for a subhalo that accounts for some of the missing complexity in the source-light distribution. At the same time, the higher level of lens modeling systematics seen in Large Masks exacerbates spurious detections. As with the source model, a subhalo can likewise account for some of the missing complexity in the lens model.
Multi-band analysis is key to disentangling source and lens modeling systematics from substructure detection. While seemingly convincing, several qualities give away hints that the perturber detected with shapelets may be spurious.
\begin{itemize}
    \item The smooth CAB parameters obtained with shapelet source modeling across different filters are in tension, putting into question the validity of the source and the lens models.
    \item Some subhalo parameters are in tension across filters when using a shapelet source model. In particular, we find an anomalously large concentration in both mask sizes in the F150W filter.
    \item The statistical significance of the subhalo detection diminishes as we complicate the source model by increasing the shapelet order parameter $n_{\text{max}}$ \rtext{or allowing for ellipticity of the shapelet basis}.
\end{itemize}

By introducing pixel-based source modeling with Delaunay triangulation, we allow for a highly flexible source model without any explicit regularization. With this higher source complexity, the preference of our model for substructure disappears. Interestingly, when imposing gradient regularization on Delaunay source modeling, spurious substructure appears again, although at different location, in two filters. This finding demonstrates that strong, non-adaptive regularization conditions can lead to spurious substructure detections, as additional complexity in the lens model accounts for the limited complexity in source modeling. This result is particularly relevant for JWST cluster lenses, where multiply-lensed, magnified images offer remarkably high resolutions, and a highly detailed source model becomes essential. However, we expect the result to translate to galaxy-galaxy strong lenses, where strong regularization conditions that limit the source model complexity could lead to a spurious preference for substructure that accounts for some of this missing complexity. The availability of multi-filter data with JWST gives us a higher chance of decisively ruling out spurious detections.

Finally, our local CAB model is in good agreement with the latest global model of Abell S1063. While this speaks for the success of novel global lens modeling techniques, we still place more trust in the ability of CAB to predict local deflections, as global models typically neglect pixel-level information due to computational challenges. To this end, summary statistics extracted from a CAB analysis can be incorporated to enhance the accuracy of global cluster models while keeping the problem computationally tractable.

\section*{Acknowledgements}
We thank Arthur Tsang for helpful discussions. CD is partially supported by the Department of Energy (DOE) Grant No. DE-SC0020223. \rtext{This work is based on observations made with the NASA/ESA/CSA James Webb Space Telescope. The data were obtained from the Mikulski Archive for Space Telescopes at the Space Telescope Science Institute, which is operated by the Association of Universities for Research in Astronomy, Inc., under NASA contract NAS 5-03127 for JWST. These observations are associated with program $\#1840$.}

%%%%%%%%%%%%%%%%%%%%%%%%%%%%%%%%%%%%%%%%%%%%%%%%%%
\section*{Data Availability}
The data is publicly available on the Mikulski Archive for Space Telescopes (MAST). 

%%%%%%%%%%%%%%%%% APPENDICES %%%%%%%%%%%%%%%%%%%%%

\appendix

%%%%%%%%%%%%%%%%%%%%%%%%%%%%%%%%%%%%%%%%%%%%%%%%%%
\section{Posteriors from Large Masks}
\label{sec:large_masks_jwst_error}
\begin{figure*}
	\centering
	\includegraphics[width=\textwidth]{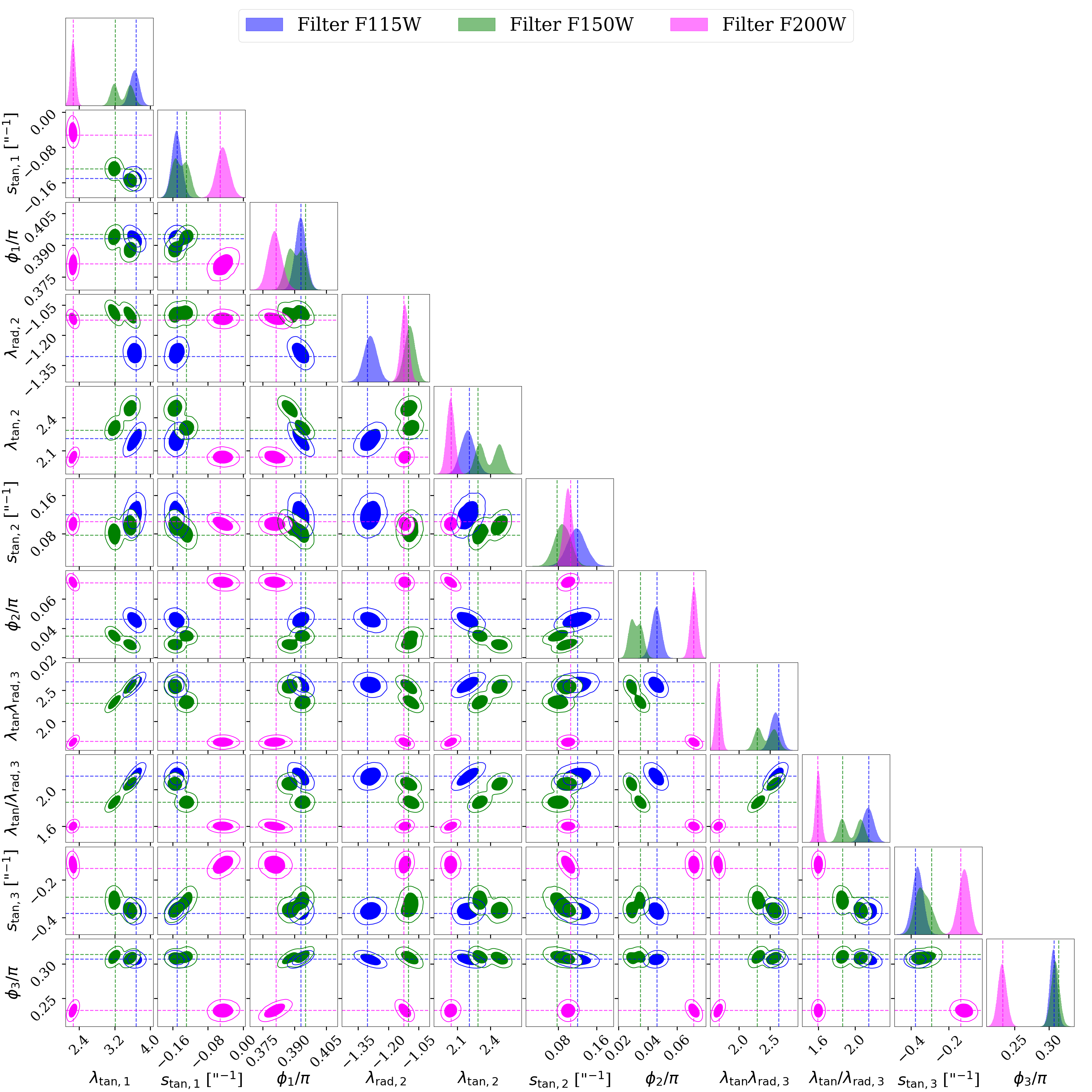}
    \caption{CAB parameters obtained using source modeling with \ptext{Cartesian shapelets} (using $n_{\mathrm{max}} = 12, 14, \text{ and } 15$ for filters F115W, F150W, and F200W, respectively) with Large Masks. The shaded and unshaded 2D contours show $1\sigma$ and $2\sigma$ confidence intervals, respectively. The dashed lines show the best-fit parameters for each filter.}
    \label{fig:cab_shapelets_original_masks}
\end{figure*}

\begin{figure*}
	\centering
	\includegraphics[width=\textwidth]{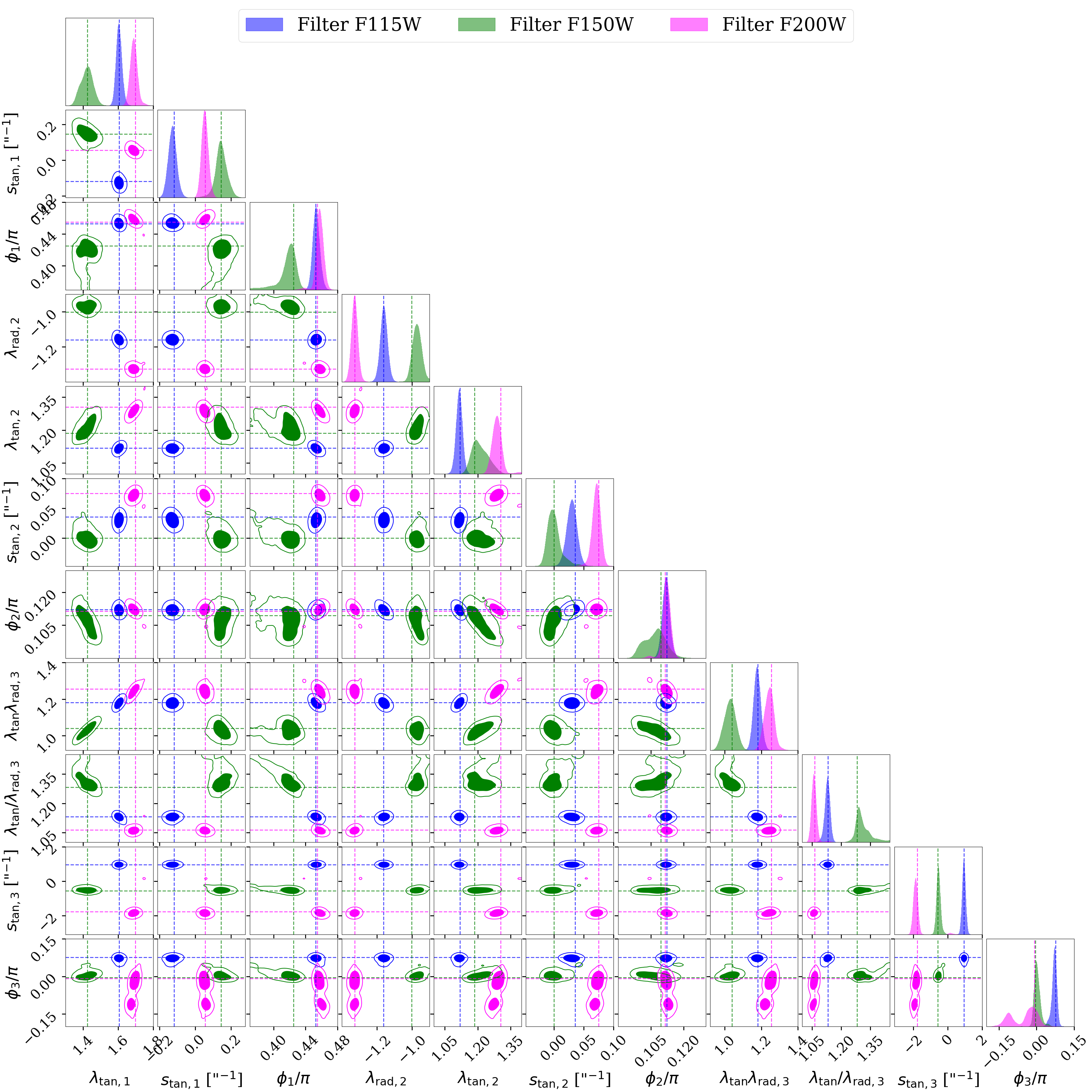}
    \caption{CAB parameters obtained with Delaunay source modeling using Large Masks. The shaded and unshaded 2D contours show $1\sigma$ and $2\sigma$ confidence intervals, respectively. The dashed lines show the best-fit parameters for each filter.}
    \label{fig:cab_delaunay_original_masks}
\end{figure*}

We show the full posterior distribution of the CAB parameters obtained with \ptext{Cartesian shapelet} and Delaunay source modeling methods, using Large Masks, in Figures \ref{fig:cab_shapelets_original_masks} and \ref{fig:cab_delaunay_original_masks}.

\section{Posteriors from Small Masks}
\label{sec:small_masks_jwst_error}
\begin{figure*}
	\centering
	\includegraphics[width=\textwidth]{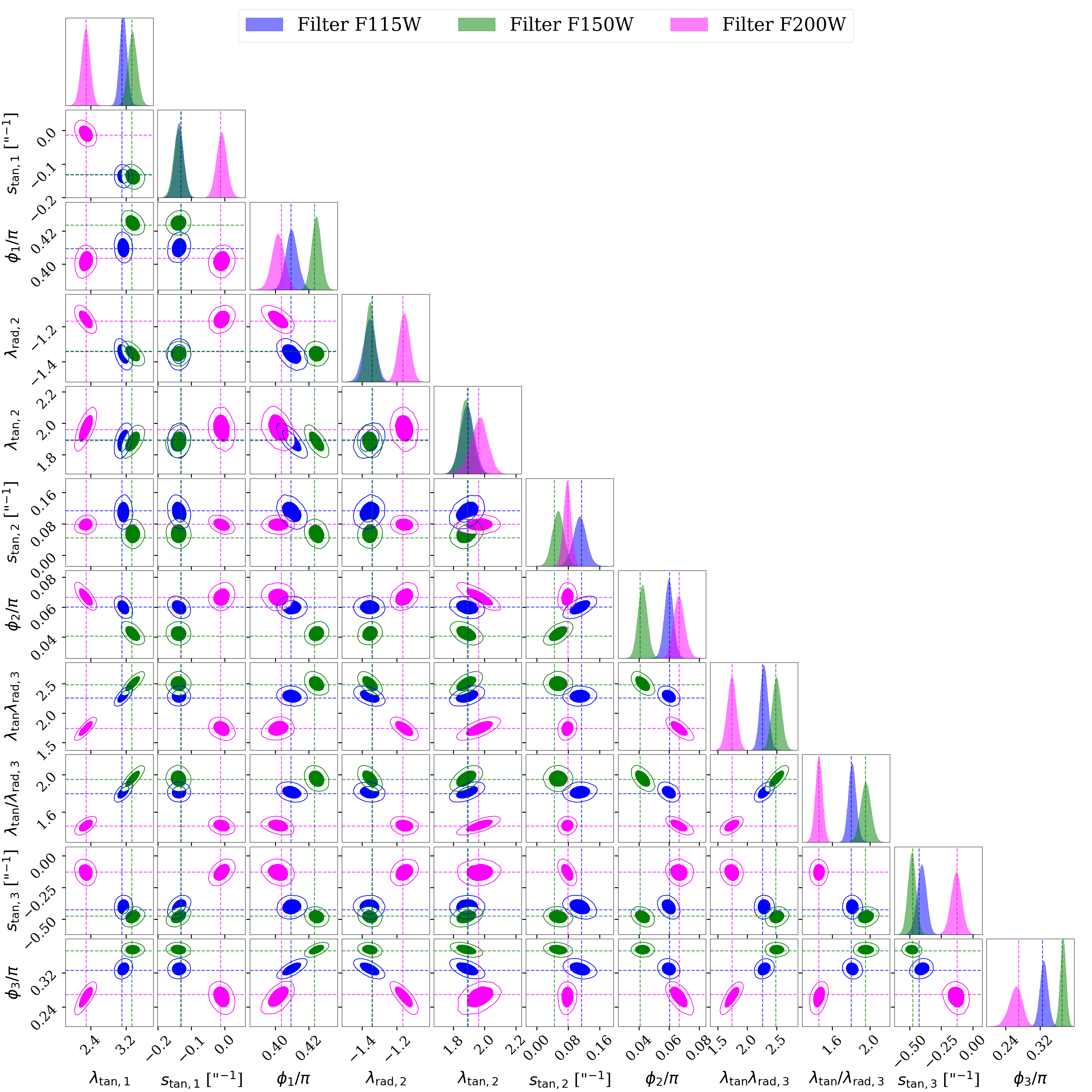}
    \caption{CAB parameters obtained using source modeling with \ptext{Cartesian shapelets} using $n_{\mathrm{max}} = 14, 13, \text{ and } 14$ for filters F115W, F150W, and F200W, respectively with Small Masks. The shaded and unshaded 2D contours show $1\sigma$ and $2\sigma$ confidence intervals, respectively. The dashed lines show the best-fit parameters for each filter.}
    \label{fig:cab_shapelets_small_mask}
\end{figure*}

\begin{figure*}
	\centering
	\includegraphics[width=\textwidth]{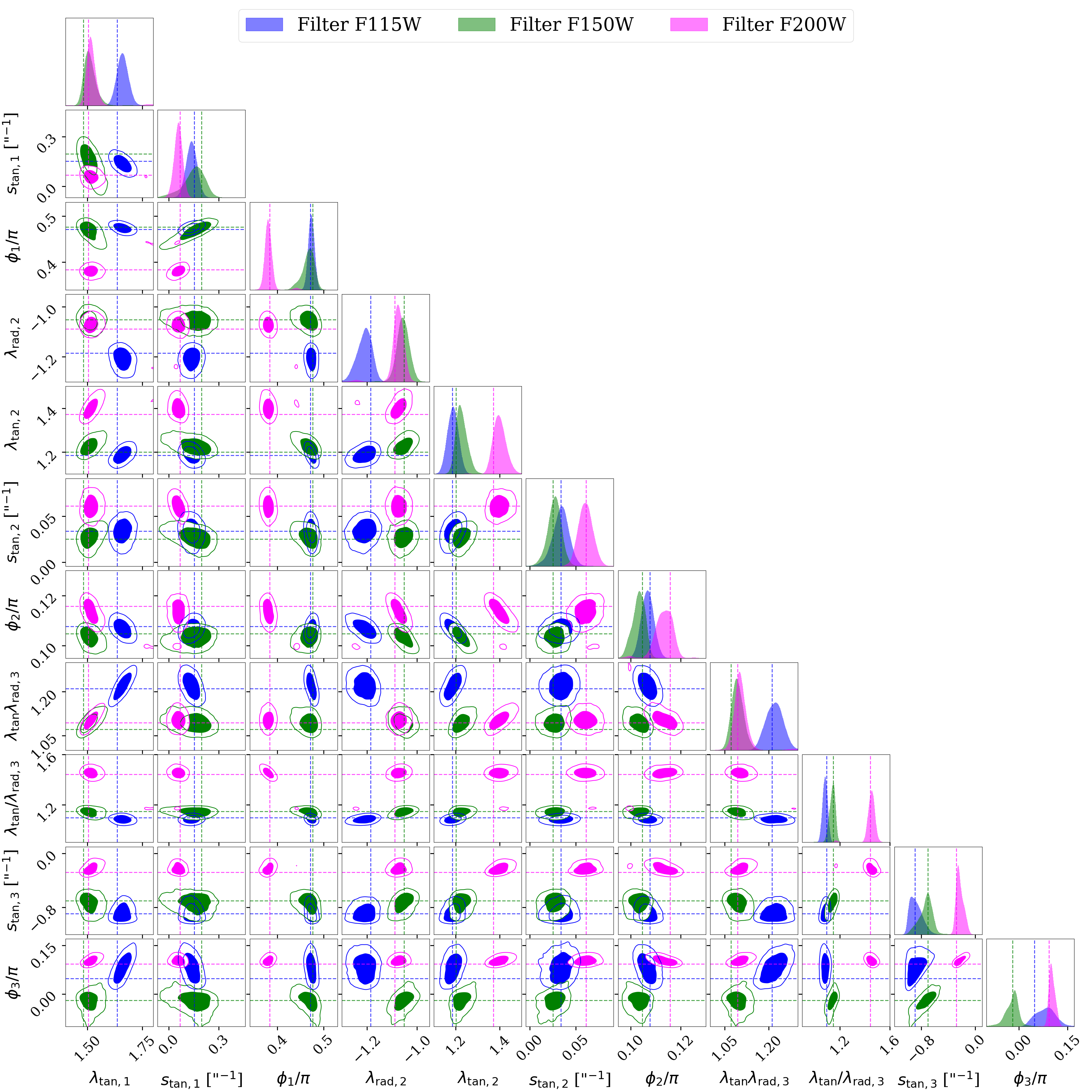}
    \caption{CAB parameters with source reconstruction with Delaunay and Small Masks. The shaded and unshaded 2D contours show $1\sigma$ and $2\sigma$ confidence intervals, respectively. The dashed lines show the best-fit parameters for each filter.}
    \label{fig:cab_delaunay_small_masks}
\end{figure*}

\begin{figure*}
	\centering
	\includegraphics[width=\textwidth]{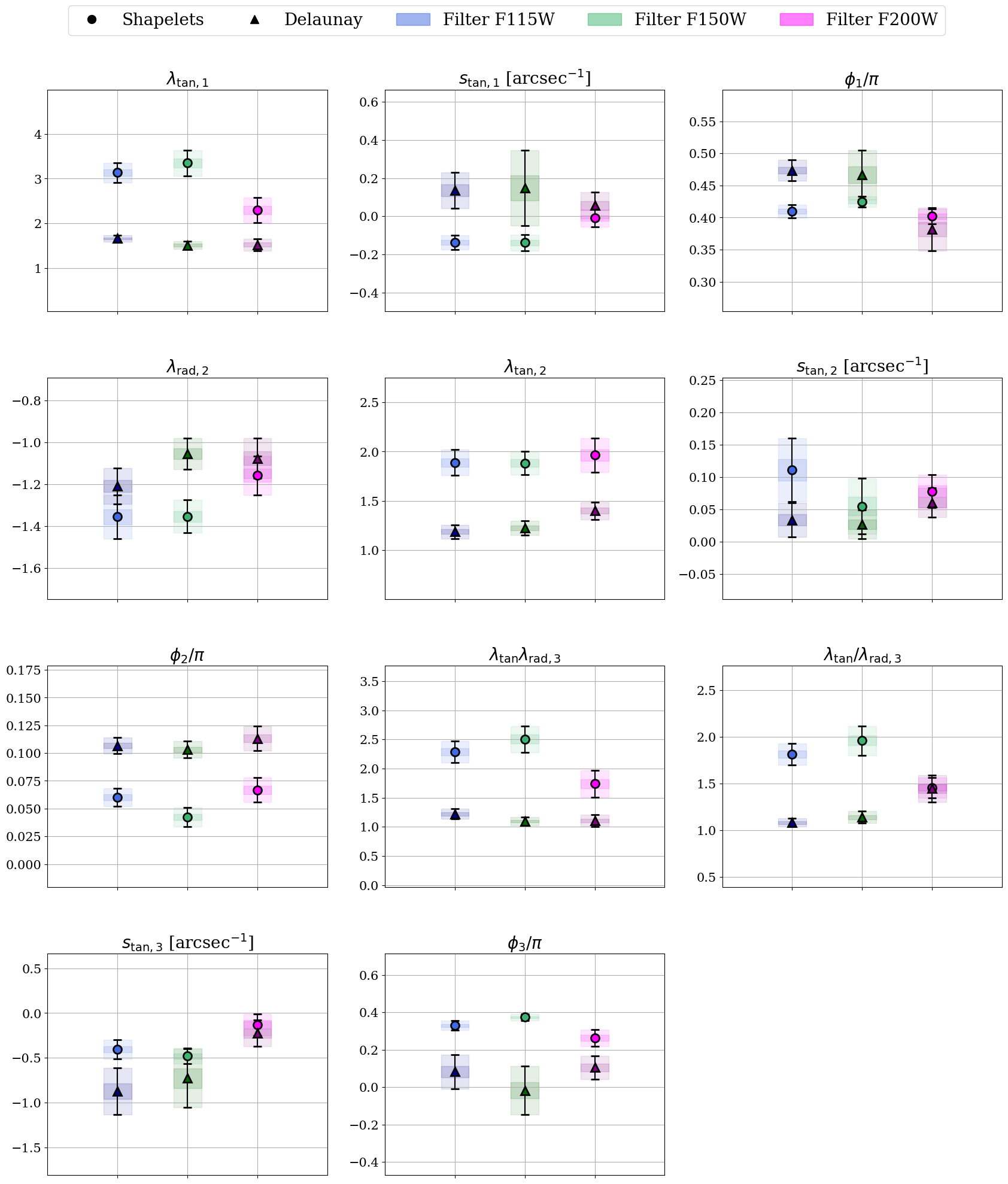}
    \caption{Means of CAB parameters obtained with \ptext{Cartesian shapelet} and Delaunay source modeling using Small Masks. Darker and lighter shaded intervals indicate the $1\sigma$ and $3\sigma$ error bars, respectively.}
    \label{fig:means_errors_small_masks}
\end{figure*}

We show the full posterior distribution of the CAB parameters obtained with \ptext{Cartesian shapelet} and Delaunay source modeling methods, using Small Masks, in Figures \ref{fig:cab_shapelets_small_mask} and \ref{fig:cab_delaunay_small_masks}. We compare the means and errors from the two source modeling methods in Figure \ref{fig:means_errors_small_masks}.

%%%%%%%%%%%%%%%%%%%%%%%%%%%%%%%%%%%%%%%%%%%%%%%%%%%%%%%%%%%%%%
\section{Spurious subhalo statistics}\label{sec:spurious_detection}
\begin{figure}
    \centering
    \includegraphics[width=0.9\linewidth]{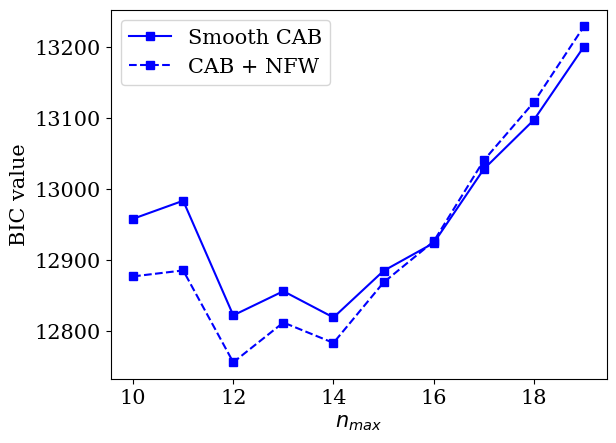}
    \caption{BIC analysis for shapelet order parameter $n_{\mathrm{max}}$ for smooth and substructure CAB models of F115W filter data (using \ptext{Cartesian shapelets and} Large Masks).}
    \label{fig:f115w_bic}
\end{figure}

\begin{figure}
    \centering
    \includegraphics[width=0.9\linewidth]{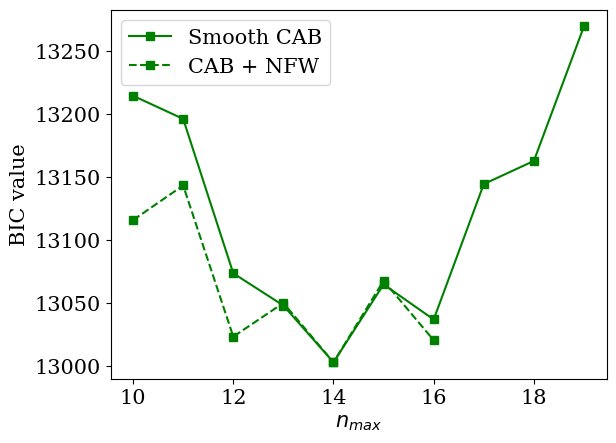}
    \caption{BIC analysis for shapelet order parameter $n_{\mathrm{max}}$ for smooth and substructure CAB models of F150W filter data (using \ptext{Cartesian shapelets and} Large Masks).}
    \label{fig:f150w_bic}
\end{figure}

\begin{figure}
    \centering
    \includegraphics[width=0.9\linewidth]{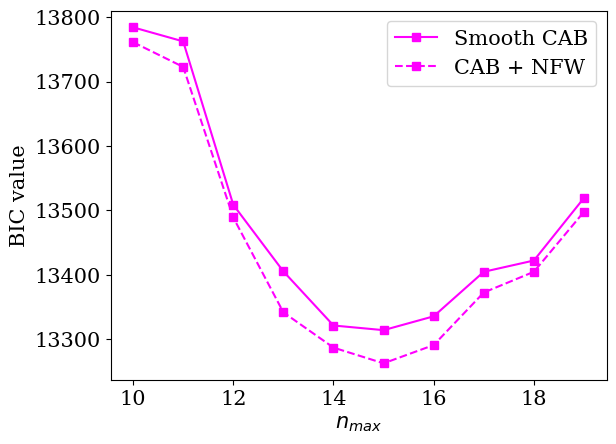}
    \caption{BIC analysis for shapelet order parameter $n_{\mathrm{max}}$ for smooth and substructure CAB models of F200W filter data (using \ptext{Cartesian shapelets and} Large Masks).}
    \label{fig:f200w_bic}
\end{figure}

\begin{figure}
    \centering
    \includegraphics[width=0.75\linewidth]{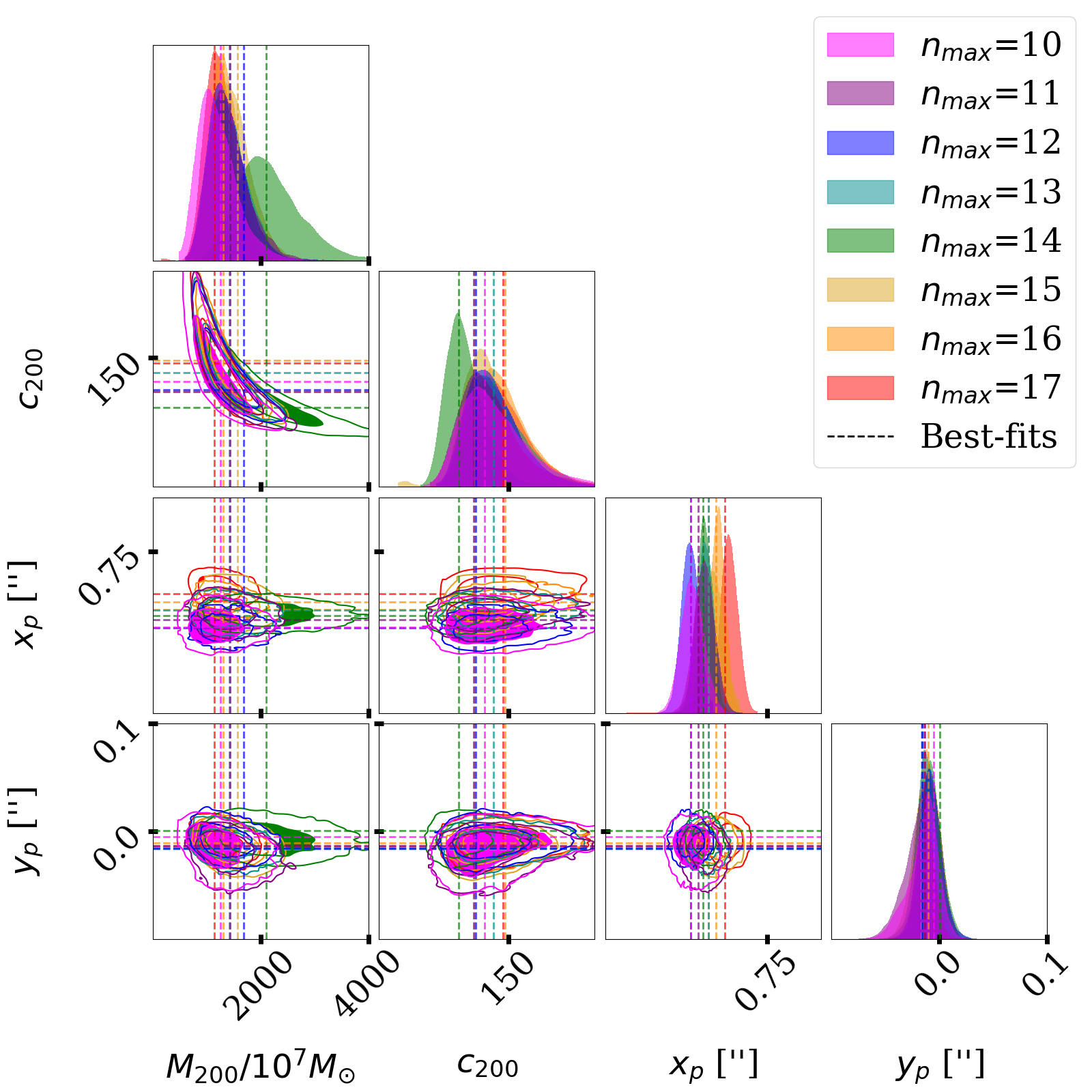}
    \caption{Posteriors of subhalo parameters in F115W with source reconstructed by \ptext{Cartesian shapelets}.}
    \label{fig:f115w_all_shapelets}
\end{figure}

\begin{figure}
    \centering
    \includegraphics[width=0.75\linewidth]{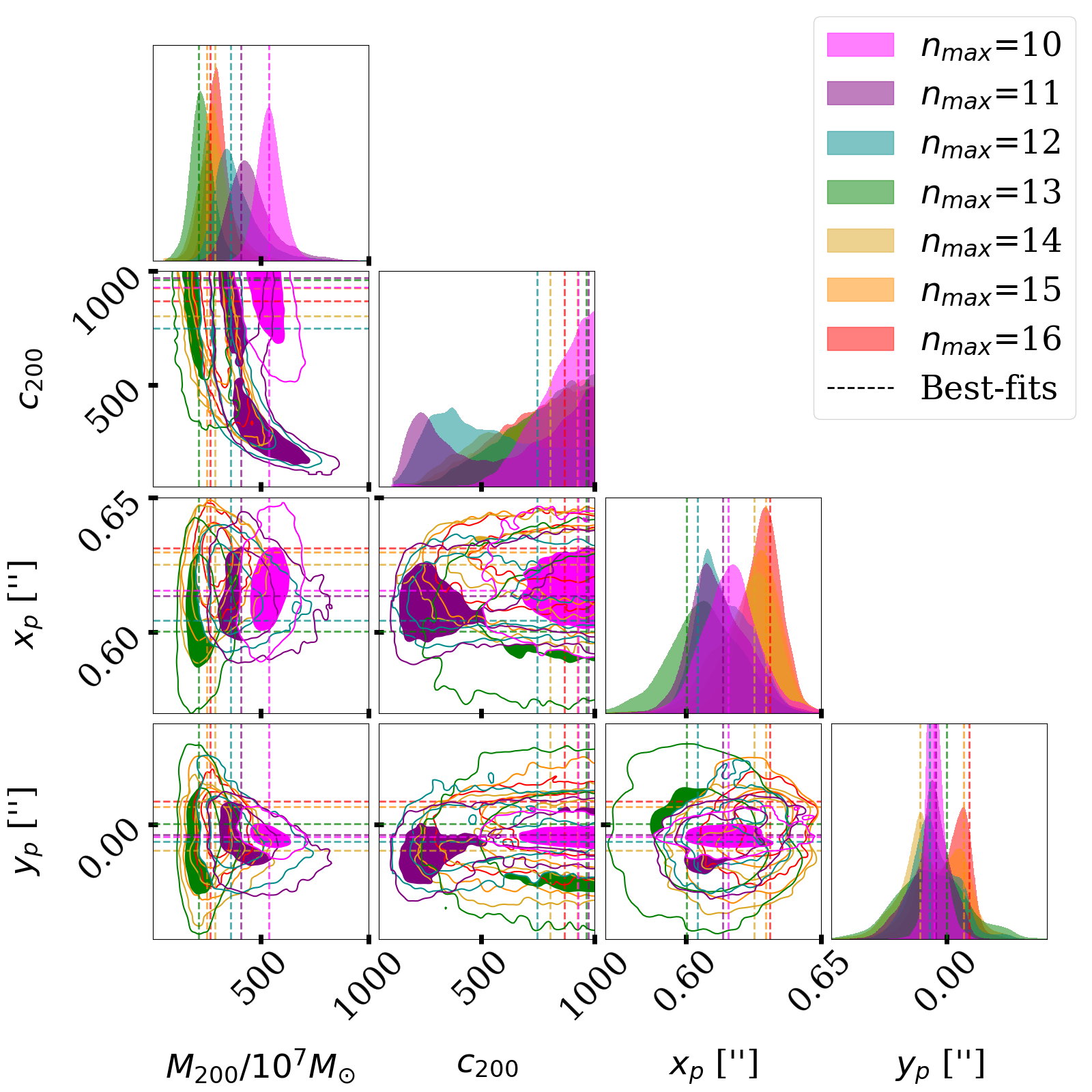}
    \caption{Posteriors of subhalo parameters in F150W with source reconstructed by \ptext{Cartesian shapelets}.}
    \label{fig:f150w_all_shapelets}
\end{figure}

\begin{figure}
    \centering
    \includegraphics[width=0.75\linewidth]{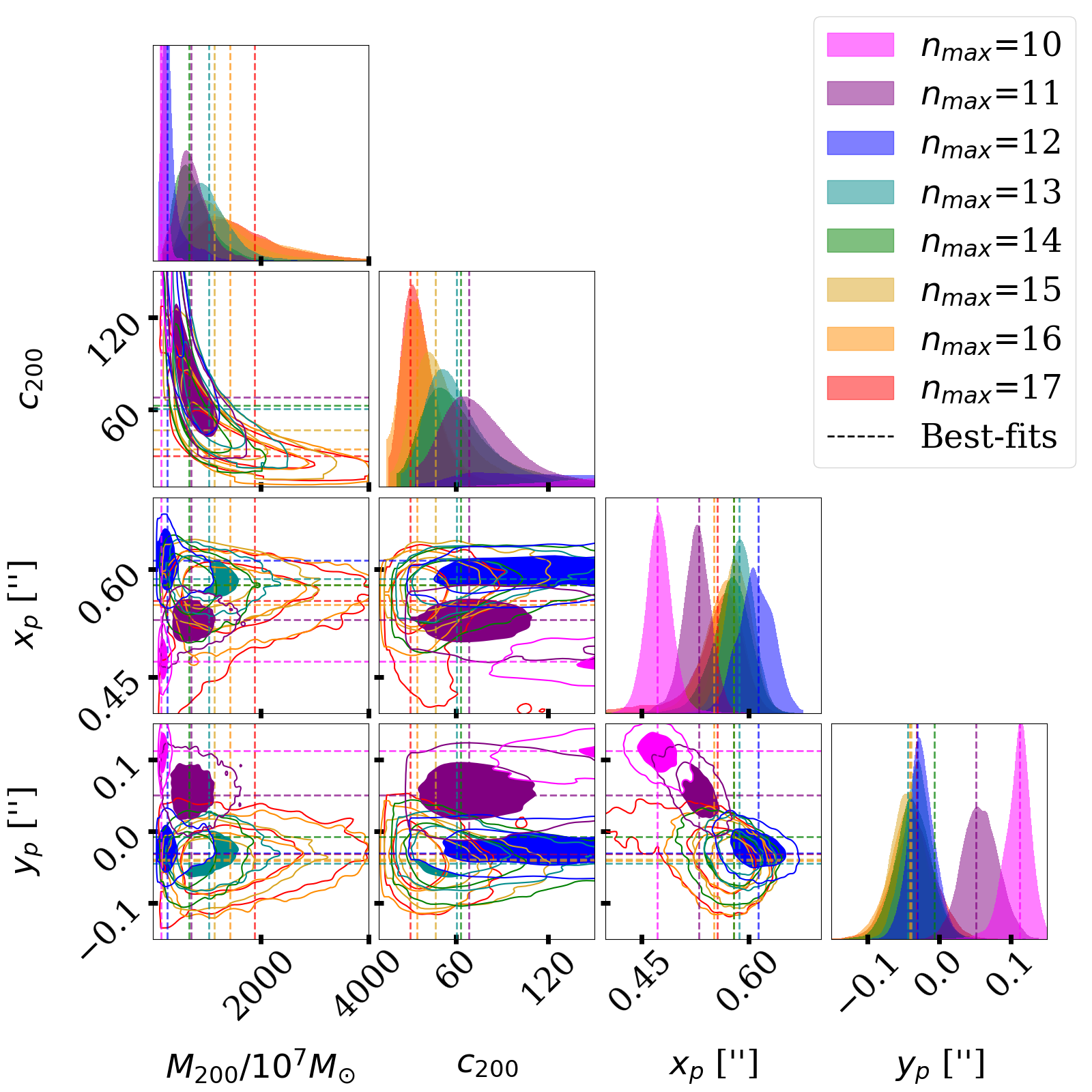}
    \caption{Posteriors of subhalo parameters in F200W with source reconstructed by \ptext{Cartesian shapelets}.}
    \label{fig:f200w_all_shapelets}
\end{figure}

We show the BIC optimization curves of the smooth CAB and substructure models in Figures \ref{fig:f115w_bic}, \ref{fig:f150w_bic}, and \ref{fig:f200w_bic}, as well as the posterior distributions of the subhalo parameters at each shapelet order parameter $n_{\mathrm{max}}$ in Figures \ref{fig:f115w_all_shapelets}, \ref{fig:f150w_all_shapelets}, and \ref{fig:f200w_all_shapelets}. Notably, the subhalo parameters are consistent across a broad range of shapelet order parameters, which shows that even more complex shapelet source models can suffer from spurious perturber detections.

%\begin{comment}

%\end{comment}

%%%%%%%%%%%%%%%%%%%%%%%%%%%%%%%%%%%%%%%%%%%%%%%%%%%%%%%%%%%%%%
\section{Elliptical Shapelets}
\label{sec:elliptical_shapelets}

\begin{figure*}
    \centering
    \includegraphics[width=\textwidth]{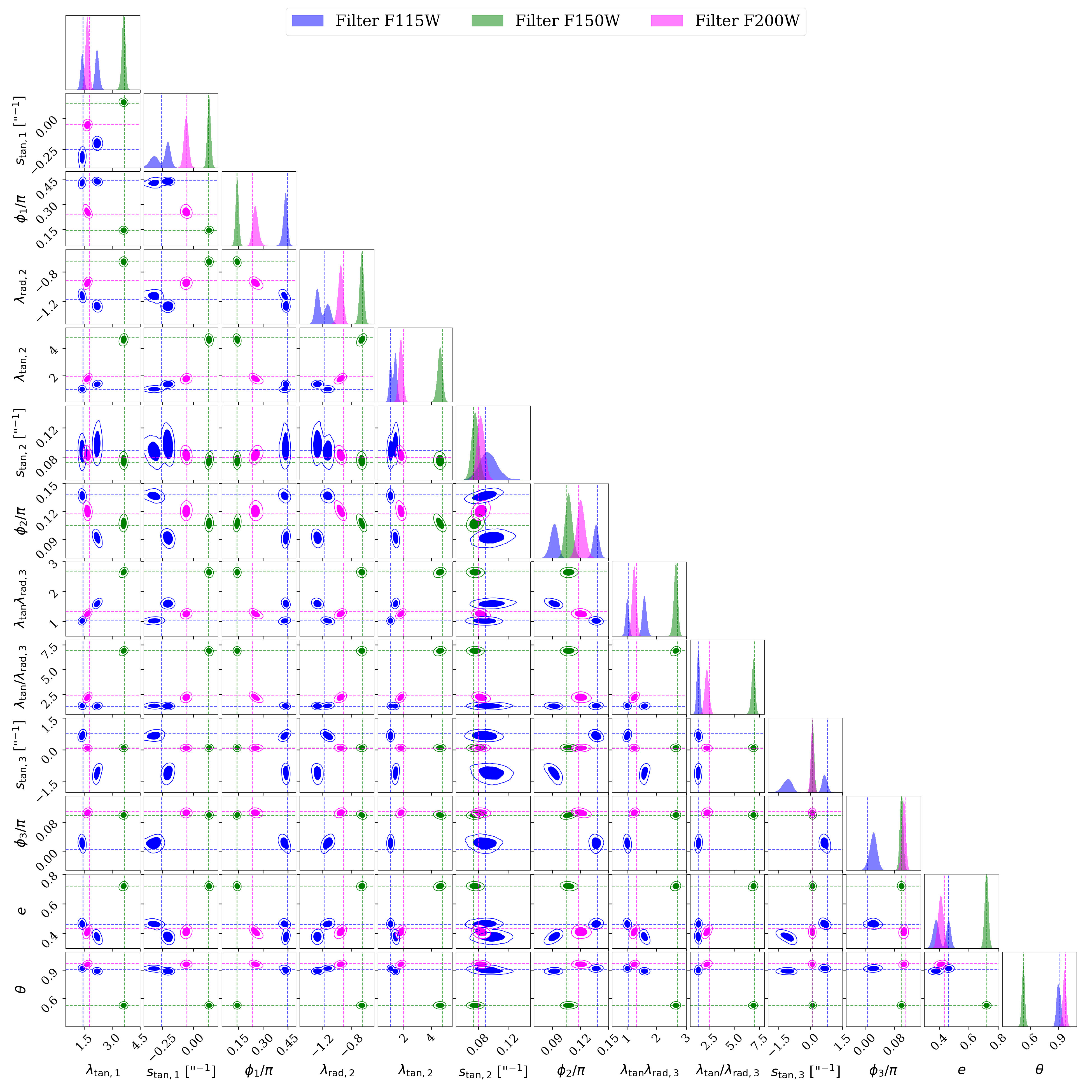}
    \caption{\rtext{Posteriors of CAB and non-linear shapelet parameters from analysis with elliptical shapelets and Large Masks at optimal shapelet-order parameters of $n_{\mathrm{max}}=12, 15, 14$ for filters F115W, F150W, and F200W, respectively.}}
    \label{fig:elliptical_large}
\end{figure*}

\begin{figure*}
    \centering
    \includegraphics[width=\textwidth]{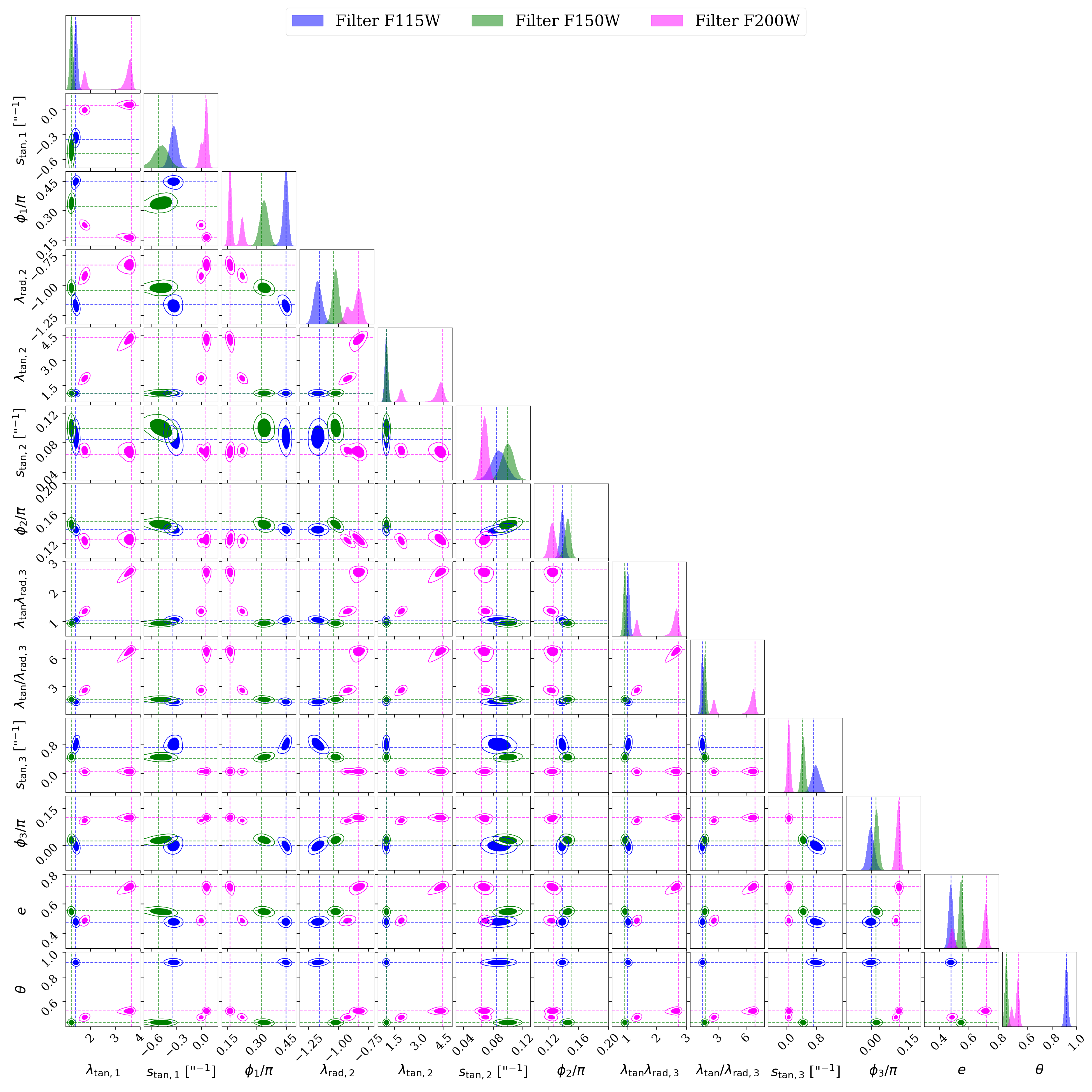}
    \caption{\rtext{Posteriors of CAB and non-linear shapelet parameters from analysis with elliptical shapelets and Small Masks at optimal shapelet-order parameters of $n_{\mathrm{max}}=12$ for all filters.}}
    \label{fig:elliptical_small}
\end{figure*}

\begin{figure}
    \centering
    \includegraphics[width=0.9\linewidth]{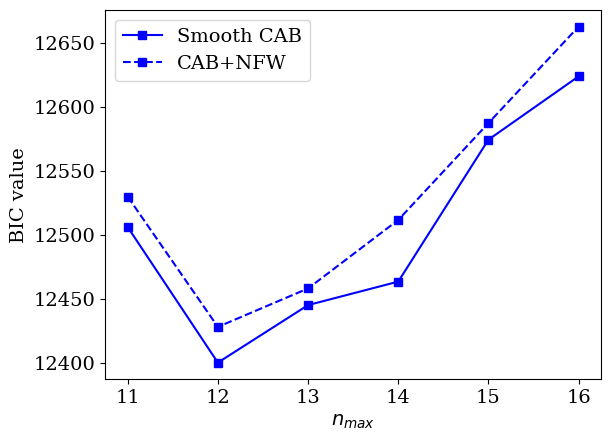}
    \caption{\rtext{BIC analysis for shapelet order parameter $n_{\mathrm{max}}$ for smooth and substructure CAB models of F150W filter data (}\ptext{using elliptical shapelets and} Large Masks).}
    \label{fig:f115w_bic_e}
\end{figure}

\begin{figure}
    \centering
    \includegraphics[width=0.9\linewidth]{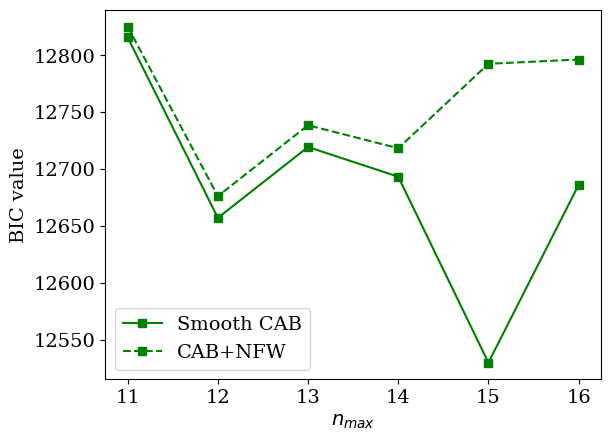}
    \caption{\rtext{BIC analysis for shapelet order parameter $n_{\mathrm{max}}$ for smooth and substructure CAB models of F150W filter data (}\ptext{using elliptical shapelets and} Large Masks).}
    \label{fig:f150w_bic_e}
\end{figure}

\begin{figure}
    \centering
    \includegraphics[width=0.9\linewidth]{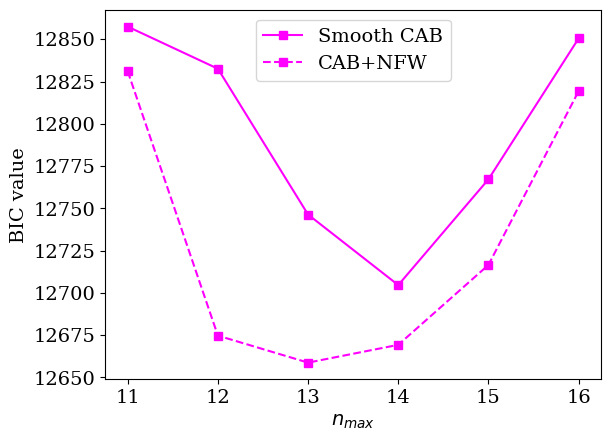}
    \caption{\rtext{BIC analysis for shapelet order parameter $n_{\mathrm{max}}$ for smooth and substructure CAB models of F150W filter data (}\ptext{using elliptical shapelets and} Large Masks).}
    \label{fig:f200w_bic_e}
\end{figure}

\begin{figure}
    \centering
    \includegraphics[width=\linewidth]{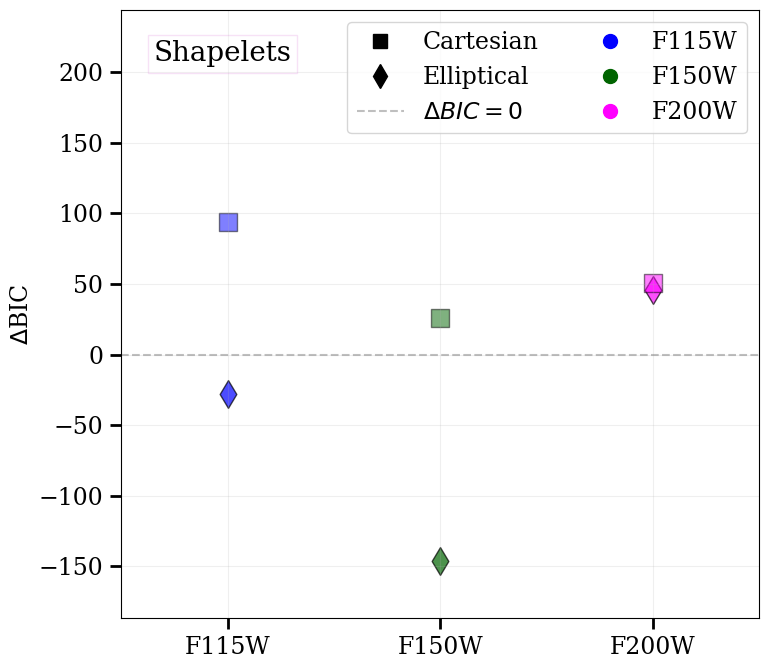}
    \caption{\rtext{$\Delta$BIC values between CAB + NFW model fits compared to smooth CAB model using Cartesian and elliptical shapelet source models in Large Masks at optimal shapelet-order parameters. All fits are performed with Wide Priors on the perturber position (see Figure \ref{fig:perturber_pos_priors}).}}
    \label{fig:dBIC_ell_large_mask}
\end{figure}

\begin{figure}
    \centering
    \includegraphics[width=\linewidth]{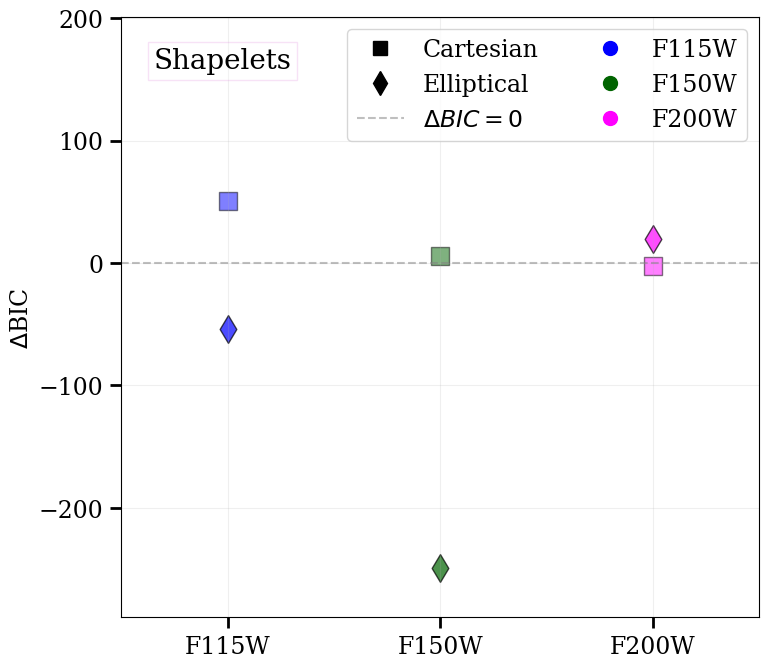}
    \caption{\rtext{$\Delta$BIC values between CAB + NFW model fits compared to smooth CAB model using Cartesian and elliptical shapelet source models in Small Masks at optimal shapelet-order parameters. All fits are performed with Wide Priors on the perturber position (see Figure \ref{fig:perturber_pos_priors}).}}
    \label{fig:dBIC_ell_small_mask}
\end{figure}

\rtext{We repeat the smooth CAB and substructure analyses using an elliptical instead of Cartesian shapelet basis. We show the resulting posterior distributions from CAB analyses in Large and Small Masks in Figures \ref{fig:elliptical_large} and \ref{fig:elliptical_small}, respectively. We show the BIC statistic of smooth and substructure CAB models for a range of shapelet-order parameters in Figures \ref{fig:f115w_bic_e}, \ref{fig:f150w_bic_e}, and \ref{fig:f200w_bic_e}. We compare the statistical significance of subhalo detection between analyses relying on Cartesian and elliptical shapelet bases for source reconstruction in Figures \ref{fig:dBIC_ell_large_mask} and \ref{fig:dBIC_ell_small_mask}.}

%%%%%%%%%%%%%%%%%%%%%%%%%%%%%%%%%%%%%%%%%%%%%%%%%%%%%%%%%%%%%%
\section{Mock Subhalo Tests}
\label{sec:mock_perturber_tests}

To compare the performance of shapelet and Delaunay source modeling methods in detecting substructures in real data, we test subhalo recovery in realistic mocks of JWST quality. We lens a real high-resolution source NGC 1300, a barred spiral galaxy, with smooth CAB deflections and a single subhalo in the line of sight of one of the lensed images. We choose a subhalo mass of $M = 10^{10} M_{\odot}$, concentration of $c = 20$, position of ($x_p$, $y_p$) = ($0.4$", $0.4$") relative to the brightest lensed image pixel, at the same subhalo redshift of $z=0.348$ as the Abell S1063 cluster. The brightness, sizes, noise levels, and Point Spread Functions of the mocks are identical to our JWST observations of \textit{system 4} in filter F115W.

We fit the mock images with \ptext{Cartesian shapelet} and Delaunay source modeling methods, using the same masks. In shapelet source modeling, we minimized the BIC of the residuals using the true CAB and NFW parameters to find the optimal shapelet order $n_{\mathrm{max}} = 25$. In Delaunay source modeling, we pick half of the unmasked pixels in the first mock image as candidate tesselation points, as described in Section \ref{sec:Delaunay_method}. If any of the chosen candidate points end up being two neighboring pixels, one of them is removed.

\begin{figure*}
    \centering
    \includegraphics[width=0.6\linewidth]{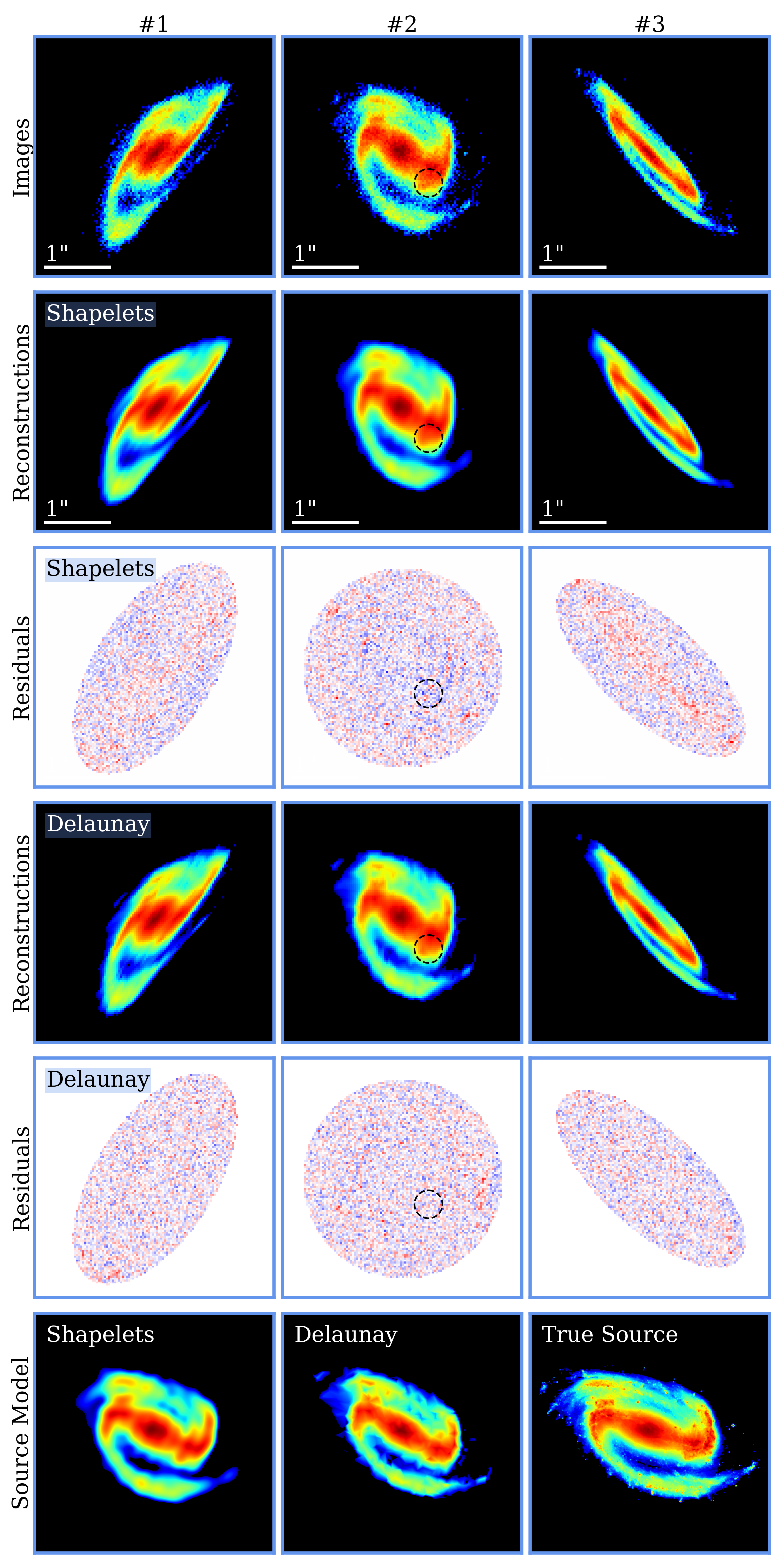}
    \caption{Model reconstructions of mock images with Delaunay and \ptext{Cartesian shapelet} source modeling methods. The shapelet model shown is with the optimal shapelet-order parameter $n_{\mathrm{max}} = 25$. The true position of the simulated subhalo is shown with a black circle.}
    \label{fig:1e10_mock}
\end{figure*}

\begin{figure*}
    \centering
    \includegraphics[width=\textwidth]{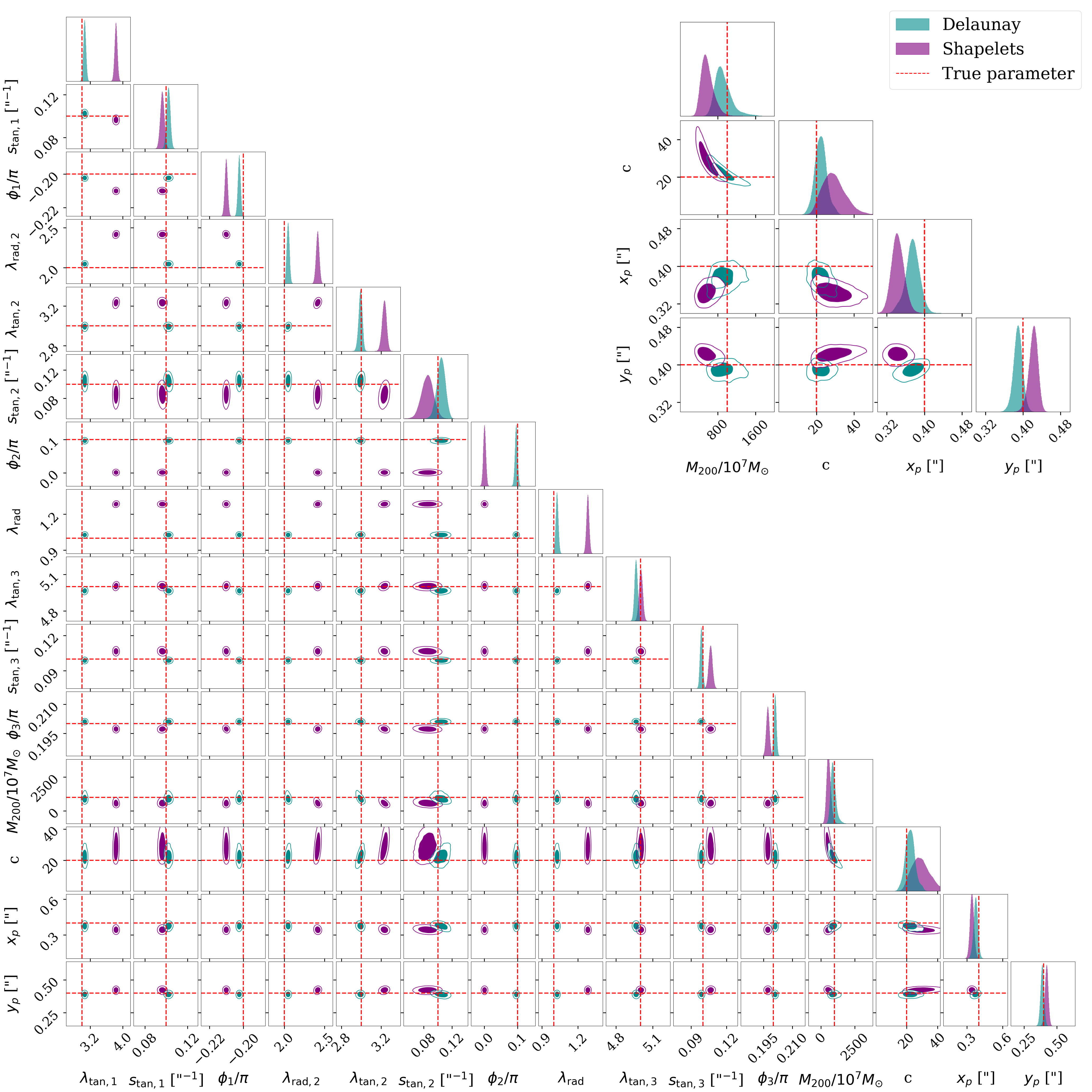}
    \caption{Posterior distribution of CAB and subhalo NFW parameters of a $M = 10^{10} M_{\odot}$ and $c = 20$ mock subhalo, measured by Delaunay and \ptext{Cartesian shapelet} source modeling methods. The shapelet fits use the optimal shapelet-order parameter $n_{\mathrm{max}} =25$, and the Delaunay fits use $N/2$ tesselation points, where $N$ is the number of unmasked pixels in the first mock image (see Fig. \ref{fig:1e10_mock}). The true parameters used to generate the mock images are shown with red dashed lines. Delaunay consistently outperforms shapelets in measuring unbiased CAB parameters, consequently leading to a more accurate measurement of the subhalo mass and concentration.}
    \label{fig:1e10_mock_fit}
\end{figure*}

We show our mock images, model reconstructions, and residuals from Delaunay and \ptext{Cartesian shapelet} source modeling in Figure \ref{fig:1e10_mock}, as well as the posterior distribution of the CAB and NFW parameters in Figure \ref{fig:1e10_mock_fit}. We find that Delaunay is consistently less biased than \ptext{Cartesian shapelet}s in measuring the smooth CAB parameters, which contributes to a more accurate measurement of the subhalo mass and concentration. We conclude that Delaunay is a reliable source modeling method for a robust substructure search, especially in lensing systems involving source galaxies with complex morphologies.

%%%%%%%%%%%%%%%%%%%% REFERENCES %%%%%%%%%%%%%%%%%%
\clearpage
\bibliographystyle{mnras}
\bibliography{main} 
\label{lastpage}
\end{document}